\documentclass[3p,times a4paper]{article}

\usepackage{amssymb}
\usepackage{wasysym}

\usepackage{lineno}

\usepackage{color,soul}
\usepackage{gensymb}

\usepackage{units}

\usepackage{subcaption}

\usepackage{hyperref}
\hypersetup{
     colorlinks   = true,
     citecolor    = gray
}

\usepackage{tocloft}

\usepackage[bordercolor=white,backgroundcolor=gray!30,linecolor=black,colorinlistoftodos]{todonotes}

\usepackage[T1]{fontenc}
\usepackage[utf8]{inputenc}
\usepackage{authblk}

\title{A hydrogen beam to characterize the ASACUSA antihydrogen hyperfine spectrometer}

\author[1,2]{C. Malbrunot \thanks{Electronic address: \texttt{chloe.m@cern.ch}; Corresponding author}}
\author[2]{M. Diermaier}
\author[2]{M.C. Simon}
\author[2]{C. Amsler}
\author[2]{S. Arguedas Cuendis\footnote{present address: Physics Department, CERN, 1211~Geneva~23, Switzerland}}
\author[3]{H. Breuker}
\author[4,5]{C. Evans}
\author[2]{M. Fleck\footnote{present address: Ulmer Fundamental Symmetries Laboratory, RIKEN, 2-1 Hirosawa, Wako, 351-0198 Saitama, Japan and Institute of Physics, University of Tokyo, 3-8-1 Komaba, Meguro-ku, 153-8902 Tokyo, Japan}}
\author[2]{B. Kolbinger}  
\author[2]{A. Lanz}
\author[4,5]{M. Leali}
\author[3]{V. Maeckel}
\author[l4,5]{V. Mascagna\footnote{present address: Dipartimento di Scienza e Alta Tecnologia, Universit\`a degli Studi dell'Insubria and INFN sez. di Pavia, Italy}}
\author[2]{O. Massiczek}  
\author[6]{Y. Matsuda}
\author[7]{Y. Nagata}
\author[2]{C. Sauerzopf}  
\author[4,5]{L. Venturelli}
\author[2]{E. Widmann}
\author[2]{M. Wiesinger\footnote{present address: Max-Planck-Institut f\"ur Kernphysik, Saupfercheckweg 1, 69117 Heidelberg, Germany}}
\author[3]{Y. Yamazaki}
\author[2]{J. Zmeskal} 
\affil[1]{European Organisation for Nuclear Research, 1211 Geneva 23, Switzerland}
\affil[2]{Stefan-Meyer Institute for subatomic physics, Boltzmanngasse 3 1090 Vienna, Austria}
\affil[3]{Ulmer Fundamental Symmetries Laboratory, RIKEN, 2-1 Hirosawa, Wako, 351-0198 Saitama, Japan}
\affil[4]{Dipartimento di Ingegneria dell'Informazione, Universit\`a degli Studi di Brescia, Brescia 25133, Italy}
\affil[5]{Istituto Nazionale di Fisica Nucleare, Sez. di Pavia, 27100 Pavia, Italy}
\affil[6]{Institute of Physics, University of Tokyo, 3-8-1 Komaba, Meguro-ku, 153-8902 Tokyo, Japan}
\affil[7]{Department of Physics, Tokyo University of Science, 1-3 Kagurazaka, Shinjuku-ku, 162-8601 Tokyo, Japan}

\begin{document}

\maketitle
\newpage
\begin{abstract}
The antihydrogen programme of the ASACUSA collaboration at the antiproton decelerator of CERN focuses on Rabi-type measurements of the ground-state hyperfine splitting of antihydrogen for a test of the combined Charge-Parity-Time symmetry. The spectroscopy apparatus consists of a microwave cavity to drive hyperfine transitions and a superconducting sextupole magnet for quantum state analysis via Stern-Gerlach separation. However, the small production rates of antihydrogen forestall comprehensive performance studies on the spectroscopy apparatus.
For this purpose a hydrogen source and detector have been developed which in conjunction with ASACUSA's hyperfine spectroscopy equipment form a complete Rabi experiment.
We report on the formation of a cooled, polarized, and time modulated beam of atomic hydrogen and its detection using a quadrupole mass spectrometer and a lock-in amplification scheme. In addition key features of ASACUSA's hyperfine spectroscopy apparatus are discussed.
\end{abstract}

\section{Introduction}
\subsection{Motivations}
\begin{figure*}
\includegraphics[trim=0cm 5cm 0cm 5cm, clip=true,width=\textwidth]{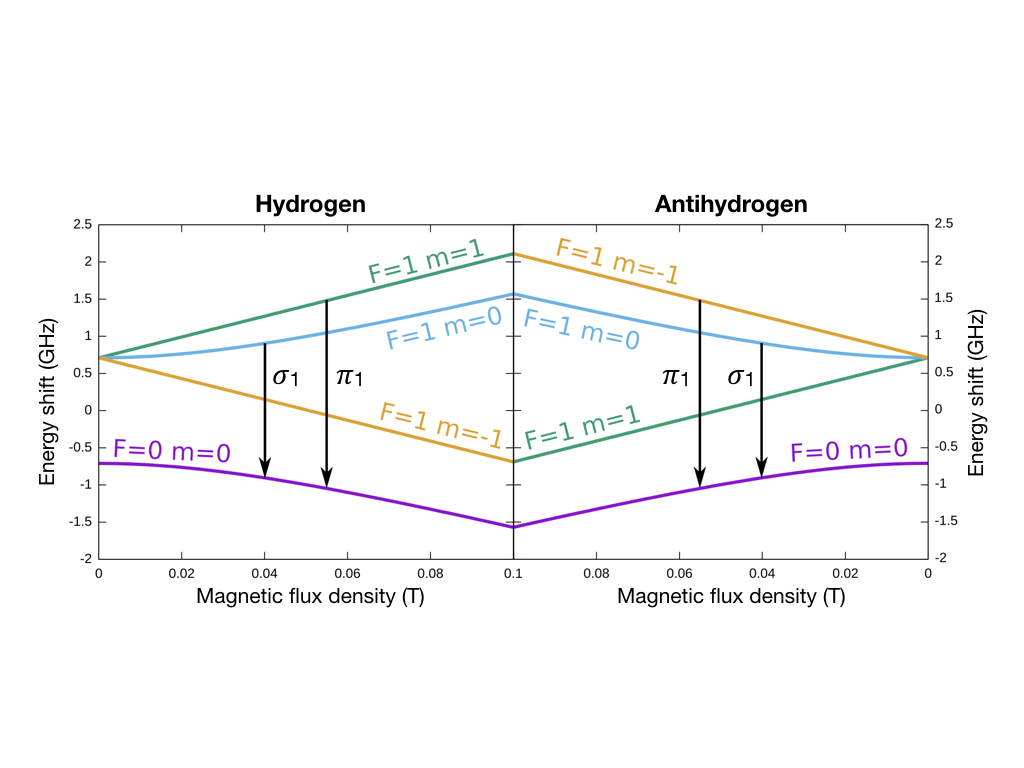}
\caption{Breit-Rabi diagram for hydrogen and antihydrogen in the presence of a small static magnetic field (modified from \cite{Clemens_diss}). The microwave transition $\sigma_1$ on hydrogen was measured with the apparatus described in this manuscript. The experimental setup to address both $\sigma$ and $\pi$ transitions will be the subject of a future publication.}
\label{fig:BreitRabi}
\end{figure*}

The hydrogen atom has motivated a plethora of experimental and theoretical investigations. Presently, a compelling reason to pursue such studies originates from the growing field of low-energy antimatter research. To date antihydrogen is the only anti-atom that can be formed in a well-controlled environment. Atomic spectroscopy methods in magnetic traps already yield precise comparisons of the hydrogen and antihydrogen spectra \cite{ALP182, ALPHA2017_2, ALPHA2018_Lyman}. Further measurements of the antihydrogen ground-state hyperfine splitting (GS-HFS) are envisioned in a beam using a Rabi-type spectroscopy apparatus described in this manuscript.  Any deviation in antihydrogen from the measured values in hydrogen would indicate a violation of  the CPT-symmetry (the combined symmetry of charge conjugation, parity, and time reversal) which would be a clear signal for physics beyond the standard model of particle physics, potentially providing new insights in the matter-antimatter asymmetry puzzle.
A measurement of the antihydrogen GS-HFS has the potential to yield one of the most precise CPT-test on an absolute energy scale. A remarkable precision of \unit[2]{mHz} (corresponding to 1.4 ppt) has been achieved in hydrogen maser experiments \cite{Hellwig, Karshenboim, Essen1, Essen2, Ramsey1, Ramsey2} owing to long interaction times by mechanical confinement. Unfortunately, this method is not transferable to antihydrogen, which would annihilate on the confining enclosure. Therefore, the method adopted within the antihydrogen program of ASACUSA (Atomic Spectroscopy And Collisions Using Slow Antiprotons), pursued by the ASACUSA-CUSP group,  at the Antiproton Decelerator (AD) of CERN is Rabi-type magnetic resonance spectroscopy \cite{Rabi1, Rabi2}. This technique requires the preparation of a polarized atomic (or molecular) beam. It is accomplished with magnetic field gradients, which result in spatial separation of atoms in different quantum states. Fig.\ref{fig:BreitRabi} shows the Breit-Rabi diagram for hydrogen and antihydrogen indicating the behavior of the atoms in an external magnetic field as a function of their magnetic moments orientations. Atoms that align their magnetic moments with the external magnetic field have a lower energy in higher fields and are called high field seekers (hfs). Atoms which follow a gradient towards lower magnetic fields are called low field seekers (lfs). 
This property allows spin state selection in so-called Stern-Gerlach type apparatus and is at the heart of the measurement technique adopted here.
Quantum transitions between lfs and hfs states are induced by means of an oscillating (or rotating) magnetic field. A second magnetic field gradient removes those atoms, that have changed their magnetic moment and the remaining ones are detected. A resonance structure can be recorded as a drop in counting rate (signal) by scanning the frequency of the oscillating magnetic field. Kusch et al.~\cite{Kusch1, Kusch2} applied Rabi-type spectroscopy to determine the ground-state hydrogen hyperfine splitting ($\sim$\unit[1.42]{GHz}) to an absolute precision of \unit[50]{Hz}. This value was improved by more than an order of magnitude using the apparatus described in the present manuscript  \cite{ourNatCommuns}.

\subsection{ASACUSA's antihydrogen hyperfine spectrometer}
 The ASACUSA-CUSP group employs a set of charged particle traps for the production of a polarized beam of antihydrogen. Antiproton bunches extracted at energies of \unit[5.3]{MeV} from the AD are further slowed down by ASACUSA's radiofrequency quadrupole decelerator \cite{BYL00} to \unit[$\sim$100]{keV} and then accumulated in a Penning-Malmberg trap (MUSASHI) \cite{musashi12}. 

\begin{figure*}[ht]
\begin{center}
\includegraphics[clip, trim=0.cm 6cm 0cm 6cm, width=1.00\textwidth]{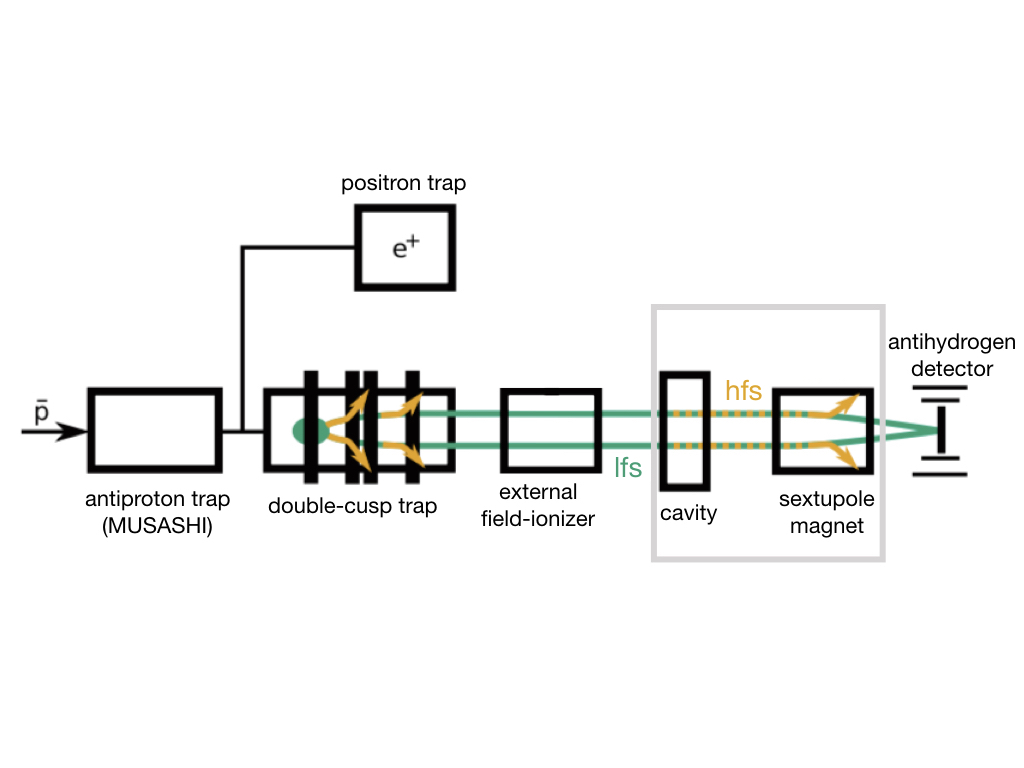}
\caption{Schematic of the ASACUSA-CUSP apparatus for the measurement of the GS-HFS of antihydrogen. The spectroscopy apparatus is indicated by the gray box.} 
\label{fig:Hbar_beamline}
\end{center}
\end{figure*}

In parallel, positrons from the $\beta^+$-decay of $^{22}\textrm{Na}$ are accumulated in a second trap. Antiproton and positron bunches are transferred to the mixing trap, which uses multi-ring electrodes and superconducting double anti-Helmholtz coils to provide both the confining electric and magnetic fields for charged particles and the magnetic field gradients for polarization of neutral antihydrogen \cite{MOH03, CUSP_Yasu, CUSP_Nagata}.
ASACUSA  reported the observation of antihydrogen atoms in a field-free environment \unit[2.7]{m} downstream of the mixing region \cite{ASACUSA_kuroda_nature_comm}, a necessary step before attempting a spectroscopy measurement. The quantum-state distribution of antihydrogen atoms exiting the formation apparatus was published \cite{Malbrunot20170273} and showed a too small fraction of ground-state atoms to achieve the spectroscopy goal. The current focus of the collaboration is thus to significantly increase the number of ground-state atoms produced.\\
\indent For the GS-HFS measurement the antihydrogen atoms extracted from the mixing trap will pass a microwave cavity for state conversion and a superconducting sextupole magnet for state analysis before being detected by an annihilation detector, as in Fig.~\ref{fig:Hbar_beamline}. 

To guarantee a large acceptance for the scarcely produced polarized ground-state antihydrogen the spectroscopy beamline has an open diameter of \unit[100]{mm}, which presents a noteworthy difference from conventional Rabi-type setups.
A cavity of the so-called strip-line geometry \cite{Kroyer, Federmann_proceedings, Silkethesis} produces the oscillating magnetic field for driving the hyperfine transitions. This resonator type was chosen, as it provides a  uniform microwave field over the large opening diameter. 
The cavity is followed by a superconducting sextupole magnet able to focus lfs ground-state antihydrogen atoms of velocities up to $\sim\unit[1000]{ms^{-1}}$ onto the antihydrogen detector, directly downstream of the sextupole.
The detector concept is based on the combination of calorimetry and track reconstruction measurements.
It comprises an inner BGO crystal read out by multi-channel PMTs providing 2D-position resolution and charge deposit information \cite{Nagata2016,NAG08}, surrounded by two layers of 32 scintillator bars each, assembled in a hodoscope geometry for tracking charged annihilation products (i.e.~mainly pions) \cite{ClemensNIMA}.
This apparatus was recently completed by two additional layers of $\unit[2\times2]{mm^2}$ square scintillating fibres perpendicular to the hodoscope bars to improve the spatial resolution in the beam direction \cite{BerniEXA}.\\ 
For characterization of the hyperfine spectrometer, the cavity and sextupole magnet, developed for the antihydrogen experiment, were coupled to an atomic hydrogen source.\\
\indent A detailed description and performance assessment of each part of the hydrogen apparatus is provided in the following sections. The source of ground-state hydrogen and its performance is detailed in \S\ref{s:chopper}, including a description of the polarization and velocity selection as well as modulation stages. The spectroscopy apparatus is described in \S\ref{s:spectro_app}: \S\ref{s:cavity} details the cavity design and performances, \S\ref{s:Heml} the Helmholtz coils' assembly for the generation of the external static magnetic field, and \S\ref{s:magn_shield} the magnetic shielding enclosing the cavity and Helmholtz coils setup while \S\ref{sec:sc6pole} provides information on the superconducting sextupole. The final descriptive section, \S\ref{s:H_detection}, deals with the hydrogen detection apparatus. Finally, \S\ref{s:characterization} provides the result of a set of characterization measurements performed with the apparatus.

\section{Hydrogen experimental apparatus}
\begin{figure*}[htp]
\begin{center}
\includegraphics[clip, trim=6.cm 0cm 5cm 0cm, width=\textwidth]{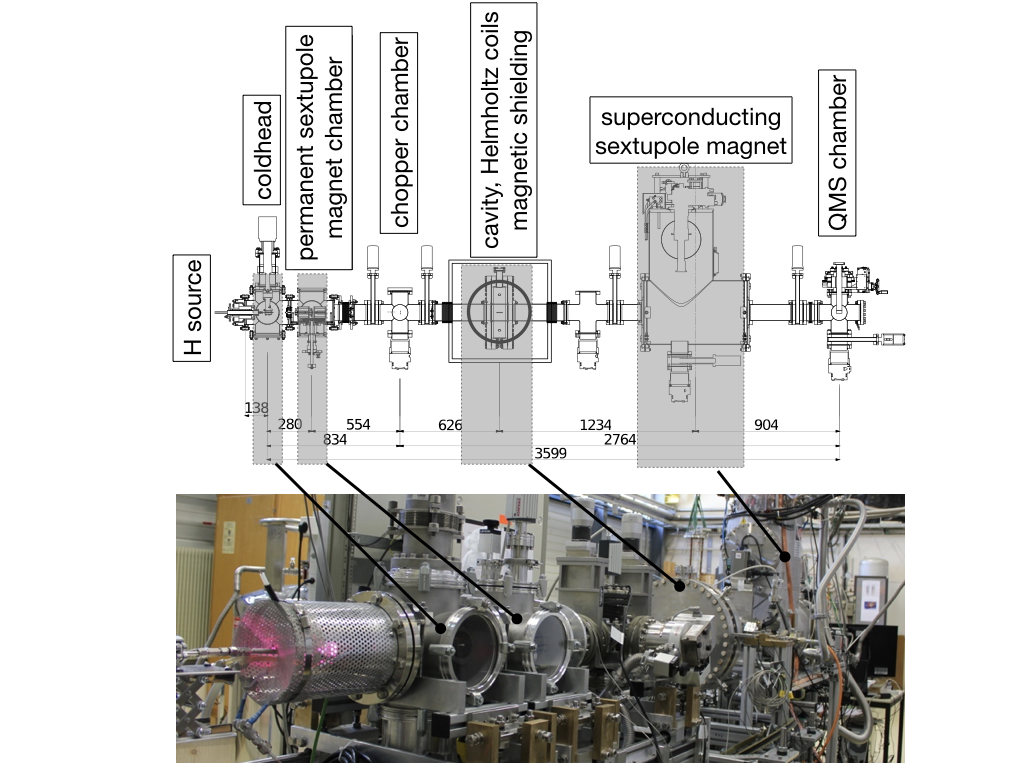}
\caption{Top : technical drawing of the atomic hydrogen beamline (dimensions in mm). Bottom : picture of the apparatus in which the shielding and Helmholtz coils around the cavity were removed. The hydrogen source is mounted directly upstream of the first vacuum chamber (the so-called straight-source configuration) and the plasma is ignited. The quadrupole mass spectrometer (QMS) stage is hidden behind the superconducting magnet.}
\label{fig:setup}
\end{center}
\end{figure*}

The hydrogen beamline is shown in Fig.~\ref{fig:setup}.
It comprised several vacuum chambers that were separated by apertures enabling differential pumping.
The first chamber enclosed the hydrogen source providing a cold ($\sim\unit[50]{K}$) beam of atomic hydrogen to the second chamber housing a set of two permanent sextupole magnets, which selected a range of velocities and polarized the beam.
The third chamber housed a chopper which modulated the beam.
The spectroscopy apparatus was placed directly downstream of the chopper chamber.
The cavity was surrounded by a pair of Helmholtz coils and enclosed in a 2-layers mu-metal box to shield the stray and earth magnetic fields.
Next, the hydrogen beam reached the bore of the superconducting sextupole magnet for spin-state analysis.
In the final chamber a quadrupole mass spectrometer (QMS) selectively counted protons (mass-1 particle), which emerged from the crossed-beam ionization region using electron impact.
The aforementioned modulation of the beam resulted in a periodical structure of the detected mass-1 rate and enabled velocity measurements from the time-of-flight and discrimination from the background via lock-in amplification. A laser shining through the entire apparatus was used for alignment as well as time-of-flight determination (see \S\ref{s:TOF}).
With the hydrogen source ignited the pressure was typically $\unit[10^{-3}]{Pa}$ in the first pumping stage and better than $\unit[5\times10^{-8}]{Pa}$ in the detection chamber.

\subsection{Polarized and modulated atomic hydrogen source}\label{s:chopper}
\begin{figure}
\centering
  \includegraphics[width=0.8\textwidth]{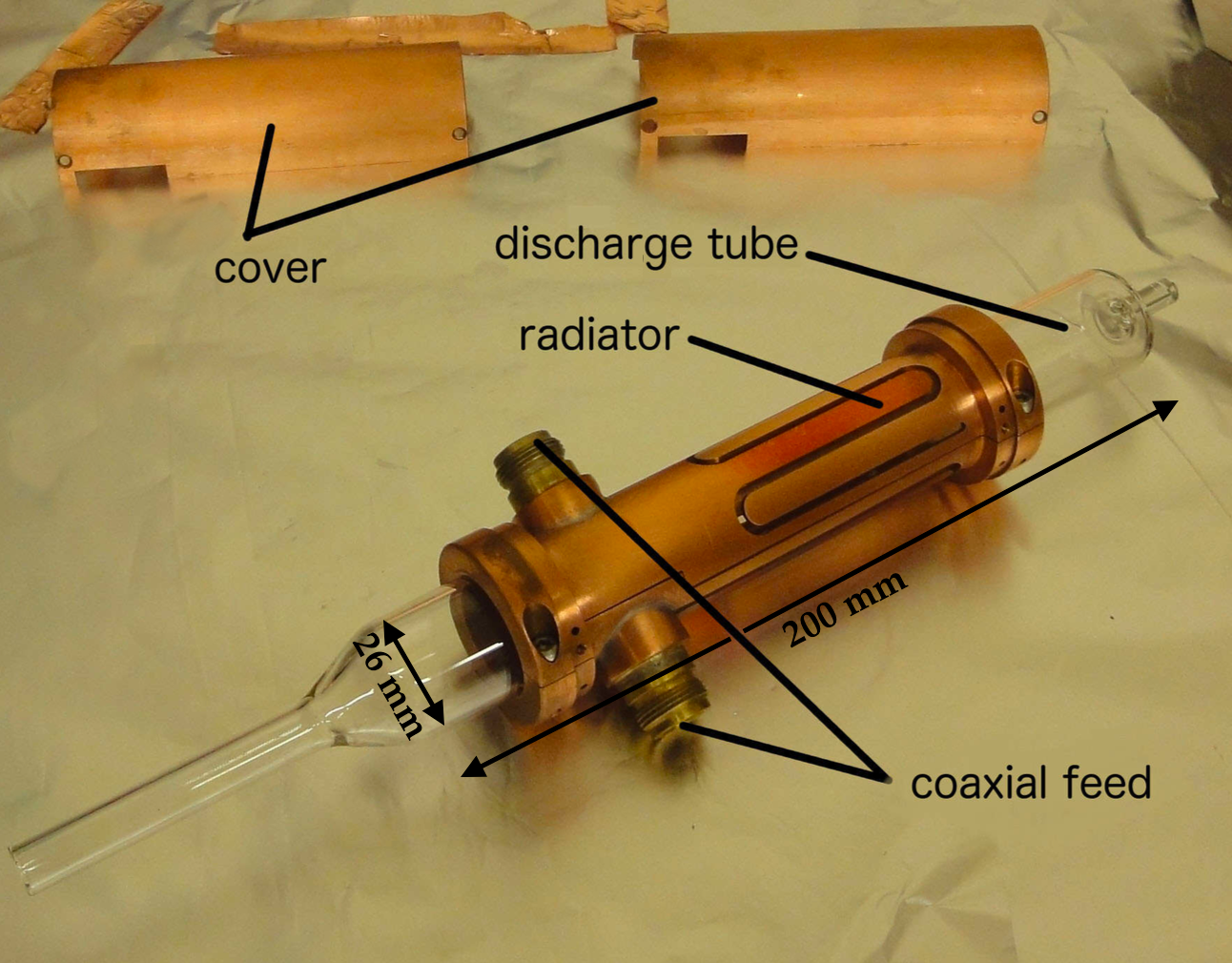}
  \caption{Photograph of the hydrogen source (modified from \cite{DIE12}). Microwaves are introduced via two N-type coaxial connectors (coaxial feeds). The slotted line antennas radiate into the glass tube and form a TE$_{011}$ mode. Molecular hydrogen is introduced at the inlet of the pyrex tube (on the left side) and atomic hydrogen exits from the right side into the vacuum system of the first chamber.}
  \label{fig:picture_source}
\end{figure}

\begin{figure}
\centering
\includegraphics[width=0.8\textwidth]{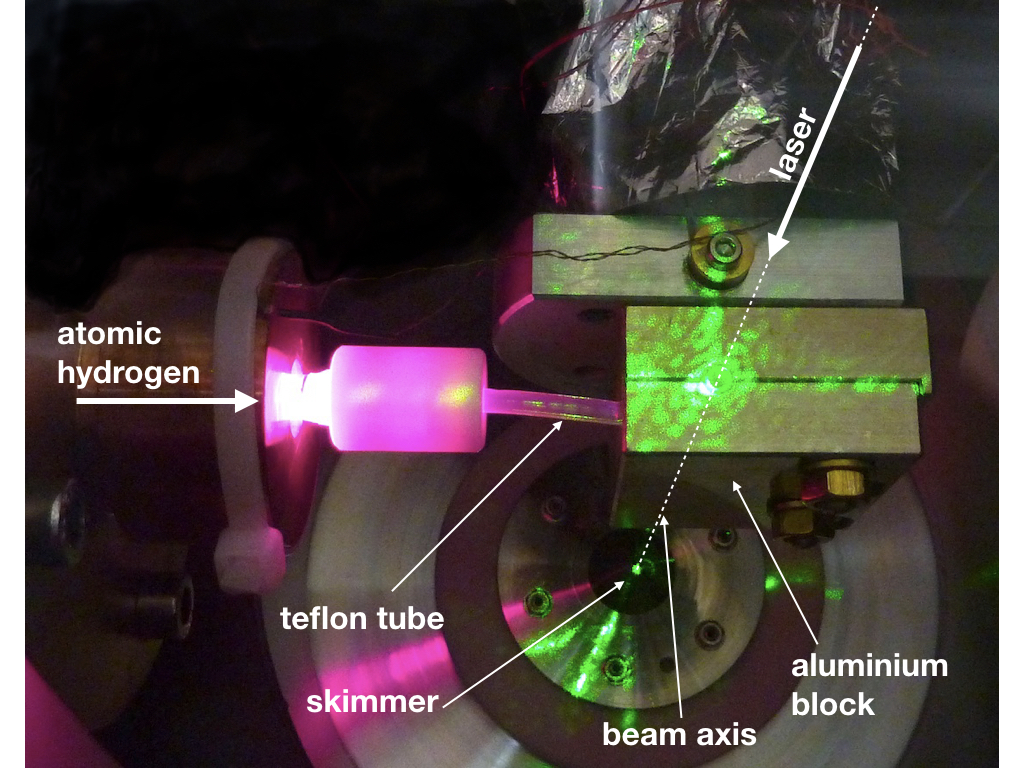}
\caption{Photograph of the source mounted with a \unit[90]{$\degree$} orientation with respect to the beam axis. Atomic hydrogen (originating from the left-side of the photograph) travel through the PTFE tubing which is bent and enclosed in aluminium blocks cooled via the cold-finger of a cryocooler. The green laser light originating from the front of the picture can be seen passing the tubing along the beam axis towards the skimmer.}
\label{fig:source}
\end{figure}

The design and operation principles of the atomic hydrogen source is similar to the one described in~\cite{McCullough}.
In the present setup, ultra pure molecular hydrogen gas was provided by a hydrogen generator (Packard 9100) via electrolysis of deionised water.
A H$_2$ flow of typically \unit[0.6]{SCCM}\footnote{SCCM is a flow unit representing a standard (at $T=\unit[273.15]{K}$ and $P=\unit[10^5]{Pa}$) cubic centimeter per minute.} was introduced into a cylindrical pyrex glass tube via an electronically controllable flow meter (Brooks SLA5850). A solid state  microwave generator (Sairem, GMS \unit[200]{W} ind C) produced $\sim$\unit[2.45]{GHz} microwaves that were fed with an input power of $\sim$\unit[60]{W} via two N-type connectors and coaxial cables into the pyrex tube.
The plasma was ignited with an electrostatic discharge gun and maintained by twin slotted Lisitano type radiators \cite{McCullough} surrounding the pyrex glass discharge tube.
A picture of the glass tube and special structure of the radiator is shown in Fig.~\ref{fig:picture_source}.
Atomic and molecular hydrogen exited through the small orifice of the pyrex tube into the vacuum system and were cooled by passing through a PTFE (Polytetrafluoroethylene) tubing kept under cryogenic temperatures ($\sim$\unit[20]{K}) by the cold-finger of a cryocooler. Two orientations of the source with respect to the beam axis were tried. The straight configuration, shown in Fig.~\ref{fig:setup}, allowed for atoms with higher velocity components traveling through the center of the PTFE tubing with minimal interactions with the walls to pass through, while a \unit[90]{$\degree$} orientation, pictured in Fig.~\ref{fig:source}, allowed for more interactions and suppressed the high velocity part of the beam.

The cracking efficiency of the hydrogen source is defined as:
\begin{equation}
D=(H_2^{\rm off}-H_2^{\rm on})/H_2^{\rm off}
\end{equation}
with $H_2^{\rm on}$ and $H_2^{\rm off}$ being the amount of molecular hydrogen in the beam when the hydrogen source is in operation or not ignited, respectively.
Operational cracking efficiencies were measured in a dedicated setup, where the source was directly connected to the detector. At a source pressure of \unit[30]{Pa} (equivalent to an incoming molecular flux of \unit[0.6]{SCCM}), maximum cracking efficiencies around 0.8 were found at room temperature which is less than what is reported in the literature (\cite{McCullough} indicates for example 0.9 at the same pressure).
Investigation gave no clear reason for this slightly lower efficiency.
The hydrogen rate was sufficient to perform the experiment and therefore no further improvements were sought.

\begin{figure}
\centering
\includegraphics[width=0.8\textwidth]{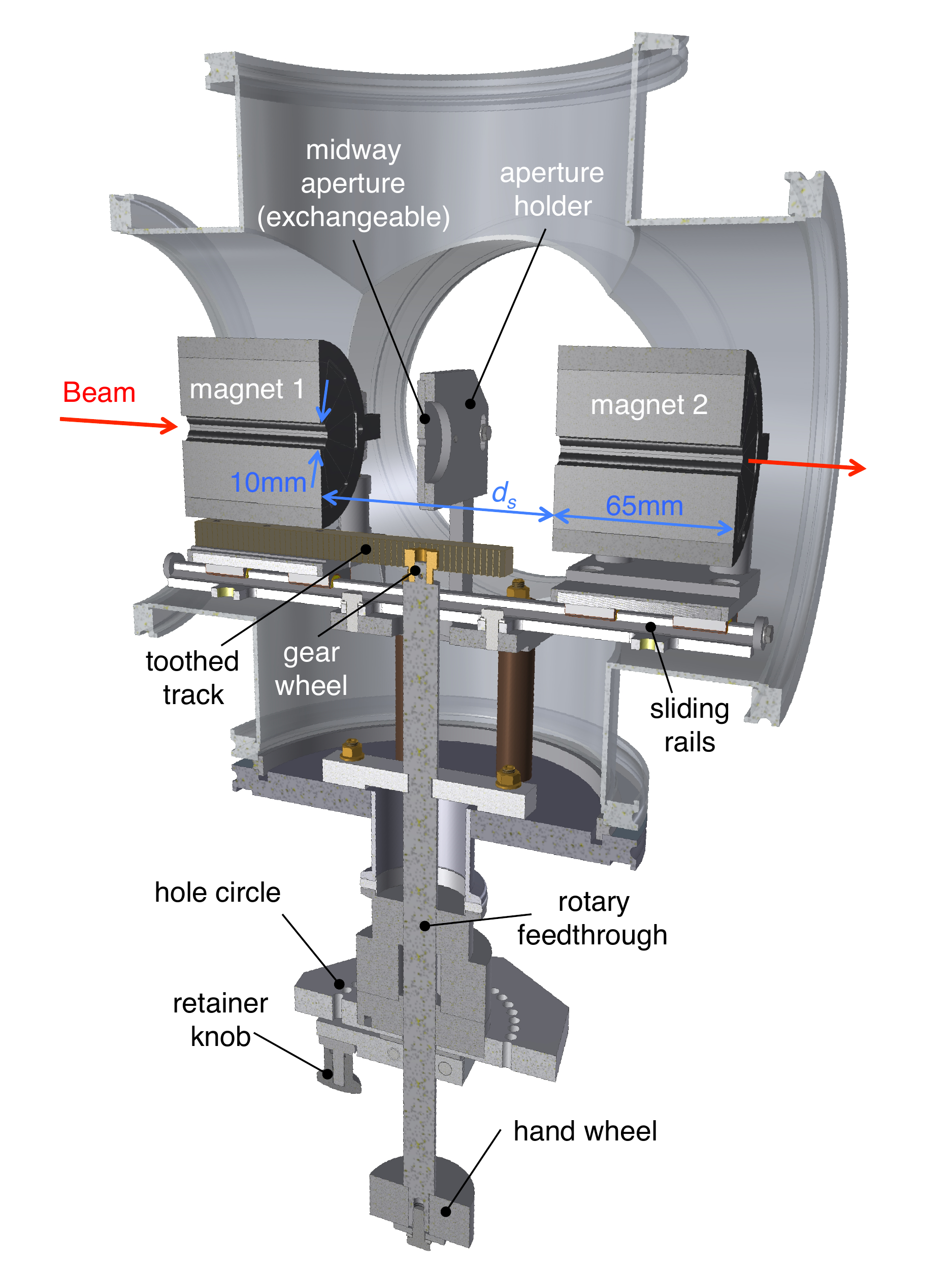}
\caption{Annotated technical drawing of the two permanent magnets assembly for beam polarization and velocity selection.}
\label{fig:sextupole_doublet_drawing}
\end{figure}

\begin{figure*}
\includegraphics[width=\textwidth]{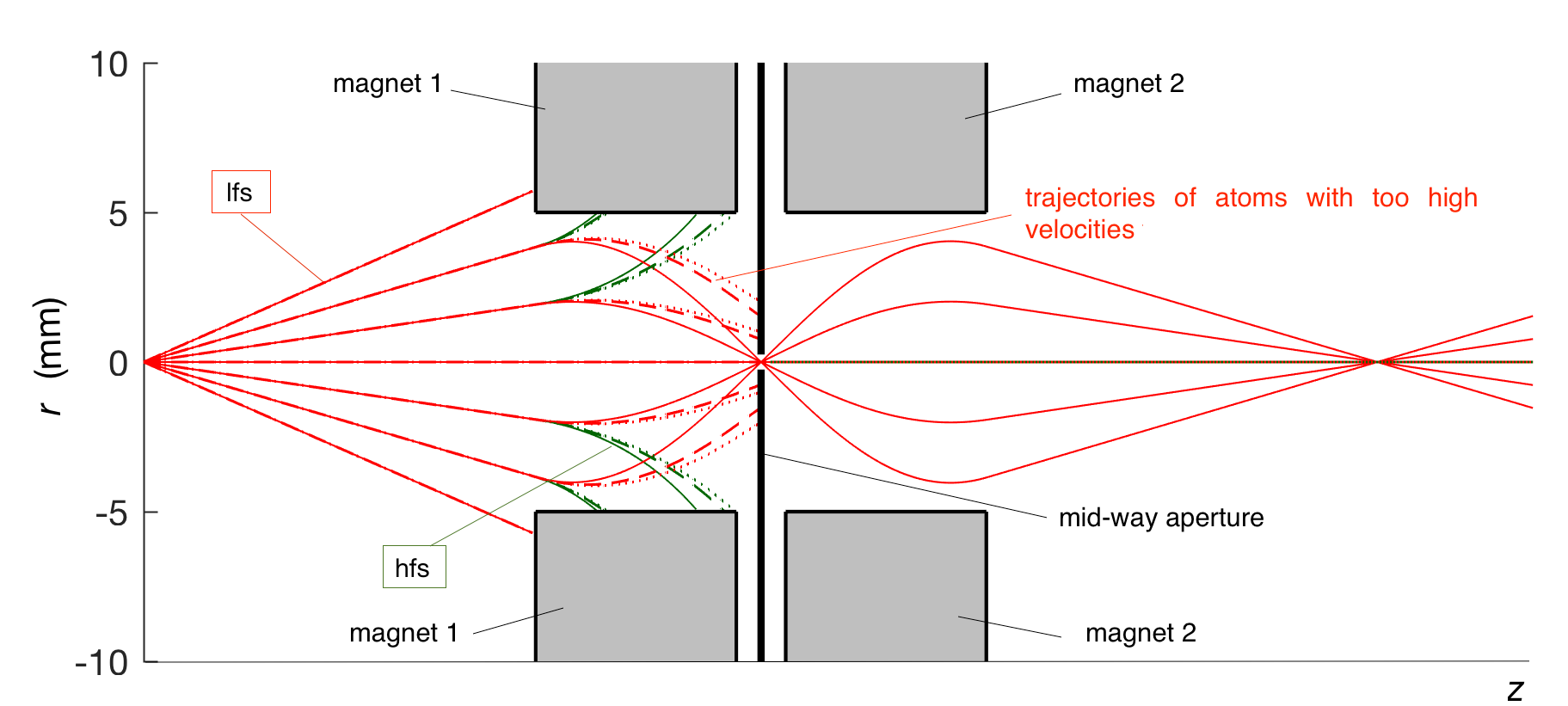}
\caption{Conceptional sketch of the velocity selection (modified from \cite{WIE16}, z-axis not to scale). Atoms in the correct spin configuration, the so-called low-field seekers (lfs, in red) will be bent onto the aperture with a velocity-dependent focal radius. Atoms in a high-seeking state (hfs, in green) will be bent away by the first magnet. The straight, dashed and dotted lines indicate the trajectories of atoms with different velocities.}
\label{fig:sextupole_doublet}
\end{figure*}

\begin{figure}
\begin{center}
\includegraphics[trim=0cm 6cm 0cm 6cm, clip=true, width=0.8\textwidth]{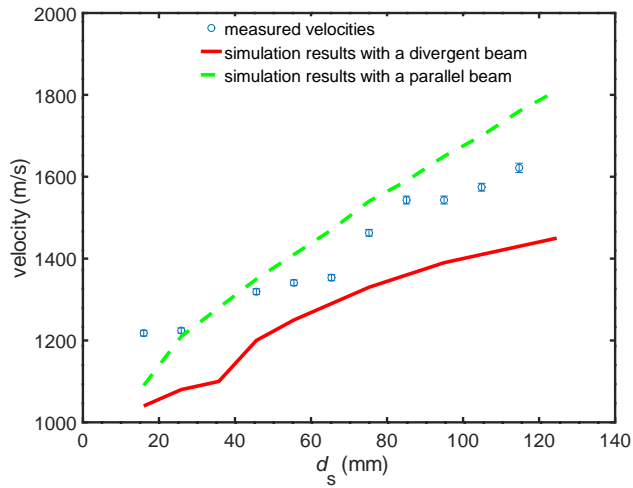}
\caption{Velocity selection as a fonction of the distance $d_{\rm s}$ between the sextupoles. The velocities were measured by time-of-flight, see \S\ref{s:TOF}. The measurements are compared to simulations where a divergent beam from the source was assumed (red solid line). In this configuration, the outgoing beam from a point-like source was assumed to be distributed within an opening angle $\alpha$ of \unit[2]{$\degree$} (200 angles simulated with $\alpha^2$ homogeneously distributed) and with initial velocities between 200 and \unit[2000]{m/s} (181 homogeneously distributed velocities were simulated). A parallel beam configuration was also simulated for comparison (green dashed line). Here, the source had a radius $r$ of \unit[5]{mm} and the atoms were originating from a homogeneously distributed $r^2$ with again 181 different velocities in the range from 200 to \unit[2000]{m/s}. For both simulations the maximum of the velocity distribution recorded after the doublet is indicated.}
\label{fig:sextupole_velocity_selection_sim_data}
\end{center}
\end{figure}

The cooled hydrogen beam exited the PTFE tubing towards a skimmer of \unit[1]{mm} in diameter and reached the second vacuum chamber hosting the set of two permanent sextupole magnets. To our knowledge, the technique to form a polarized and velocity-selected beam using a sextupole magnet doublet has not been discussed in the literature. Some details on the setup and working properties will therefore be provided in the following.
The magnets had an open diameter of \unit[10]{mm} and a pole field of $\sim$\unit[1.36]{T}, which corresponds to a gradient-constant of $G'$=\unit[108,800]{Tm$^{-2}$} (a definition of this strength and more details on the properties of sextupole fields are given in subsection \ref{sec:sc6pole}, in context with the superconducting analysis magnet).
With an mechanical length of \unit[65]{mm}, an integrated gradient (focusing strength) of \unit[7,072]{Tm$^{-1}$} can be estimated, in reasonable agreement with the design value of \unit[7,435]{Tm$^{-1}$} \cite{Sextupole_CERN}.
The magnets were made of iron-dominated poles (permendur) and samarium-cobalt Sm$_2$Co$_{17}$ permanent magnet blocks acting as magnetic flux generator enclosed in a non-magnetic titanium frame.
The first magnet provided the initial spin-polarization of the beam by removing the two hyperfine states which were attracted to high fields (hfs).
In conjunction with the second magnet and an aperture of \unit[3]{mm} in diameter between the two magnets a narrow velocity range was selected. The magnet's support mechanism governed a symmetric longitudinal motion of the two sextupoles with respect to the midway aperture, see Fig.~\ref{fig:sextupole_doublet_drawing}.
Since the focusing length depends on the atom velocity, only a given velocity component, depending on the distance between the sextupole magnets ($d_\textrm{s}$),  was focused onto the aperture and went through, as illustrated in Fig.~\ref{fig:sextupole_doublet}.\\
\indent The velocity-selected beam entered the third vacuum chamber through an aperture of \unit[5]{mm} in diameter.
In this chamber, a tuning fork chopper (Scitec CH-10) modulated the beam. It operated with a maximal opening of \unit[5]{mm}, a duty cycle of \unit[50]{\%} and a fixed frequency of $\sim$\unit[178]{Hz}.
The modulation enabled a statistical measurement of the time-of-flight (TOF) of the atoms from the chopper to the detector, by comparison of the detector signal to a sinusoidal reference signal from the chopper driver. The operation principle and chopper design originated from \cite{WAL82} which pioneered the development of helium-temperature hydrogen sources.\\
Figure~\ref{fig:sextupole_velocity_selection_sim_data} shows simulations of the sextupoles' velocity selection compared to experimental results based on TOF measurements (see \S\ref{s:TOF}) for the \unit[90]{$\degree$} source orientation. The measurements were done on a dedicated setup where the cavity and the downstream superconducting sextupole were removed and the latter replaced by a set a permanent sextupoles, installed directly after the chopper chamber, to focus the beam. The distances between the source and the sextupole doublet and the chopper were however identical to the ones in Fig.~\ref{fig:setup}. The distance between the chopper and the QMS detector in the configuration used for this velocity characterization was \unit[2.2]{m} (about \unit[0.5]{m} less than in the presently described setup, see Fig.~\ref{fig:setup}). The measured data lie between the two assumed initial beam distributions in the simulation and the trend is well reproduced. \\
\indent The achieved level of polarization of the beam by this magnet assembly has been studied in simulations which indicated a proportion of low-field seeking states ($\frac{\textrm{lfs}}{\textrm{lfs}+\textrm{hfs}}$) of more than \unit[90]{\%}.
The experimental determination of the polarization is hindered in the herein described apparatus because, apart from a possible hfs contamination, only one of the two low-field seeking states present in the beam can be addressed with the $\sigma_1$ transition, see Fig.\ref{fig:BreitRabi}.
An upgraded apparatus able to address the content of both lfs states (through the $\sigma_1$ and $\pi_1$ transitions) and therefore determine the polarization of the beam will be described in a future publication.

\subsection{Spectroscopy apparatus}\label{s:spectro_app}

The spectroscopy apparatus, composed of the microwave cavity inducing transitions between the internal states of ground state hydrogen, and the superconducting sextupole magnet analyzing the spin state, was separated from the chopper chamber by an additional aperture.
Two types of apertures were used:~the first type had a wall thickness of \unit[3]{mm} and an open diameter of \unit[4]{mm}, producing a narrow beam of approximately \unit[8]{mm} in diameter at the cavity centre.
The second aperture type, to increase the beam diameter at the cavity for systematic studies, while retaining a good differential pumping, was made of a \unit[100]{mm} long pipe with a diameter of \unit[15]{mm}, producing a beam with a diameter of approximately \unit[22]{mm} in the cavity. 

\subsubsection{Cavity} 
\label{s:cavity}
\begin{figure}
\begin{center}
\includegraphics[trim=8cm 6cm 6cm 8cm, clip=true, width=0.8\textwidth]{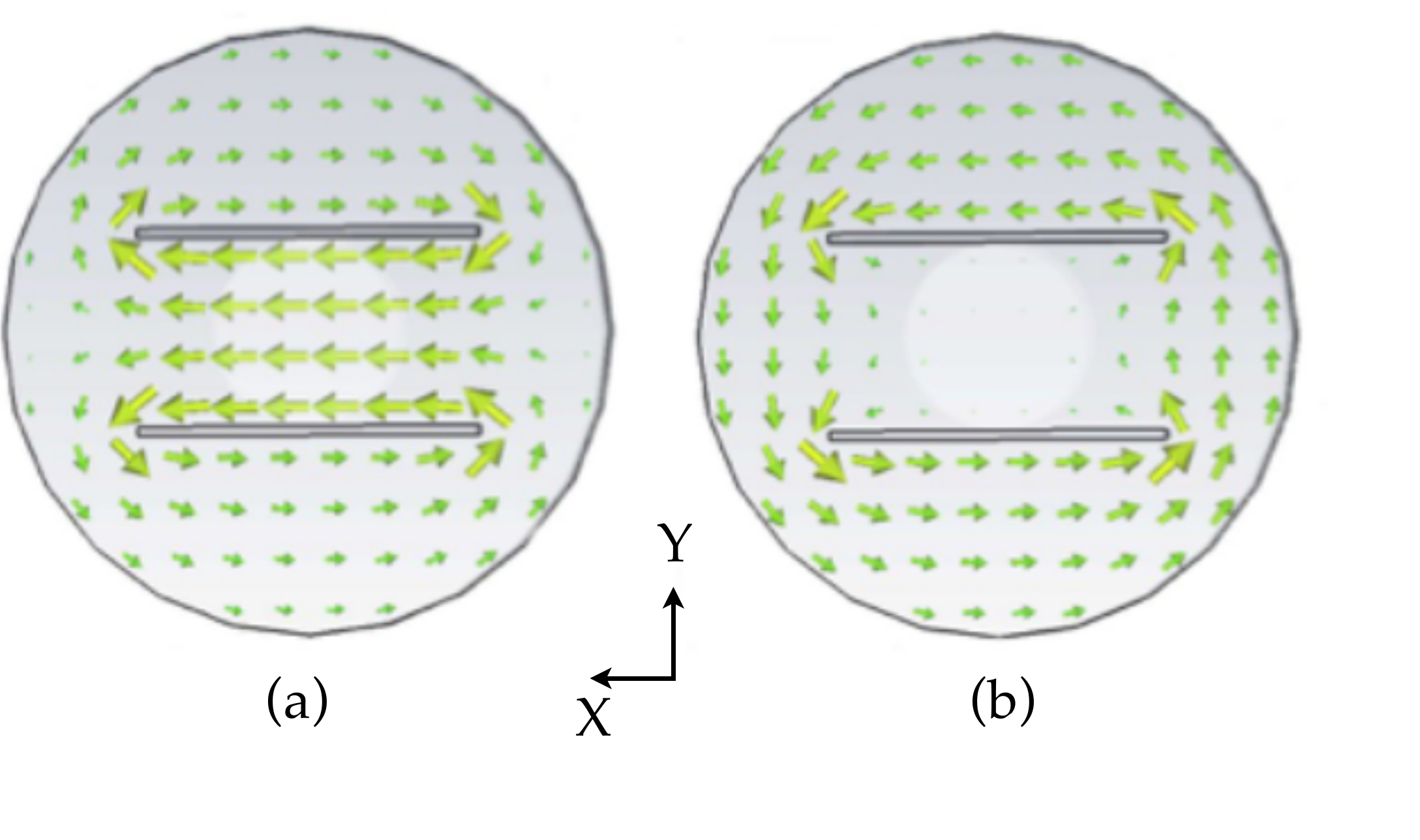}
\caption{Magnetic field distribution of the desired odd (a) and the undesired even mode (b) generated inside the cavity (modified from \cite{Silkethesis}).}
\label{fig:oddmode}
\end{center}
\end{figure}

\begin{figure}
\begin{center}
\includegraphics[trim=8cm 0cm 4cm 0cm, clip=true, width=0.8\textwidth]{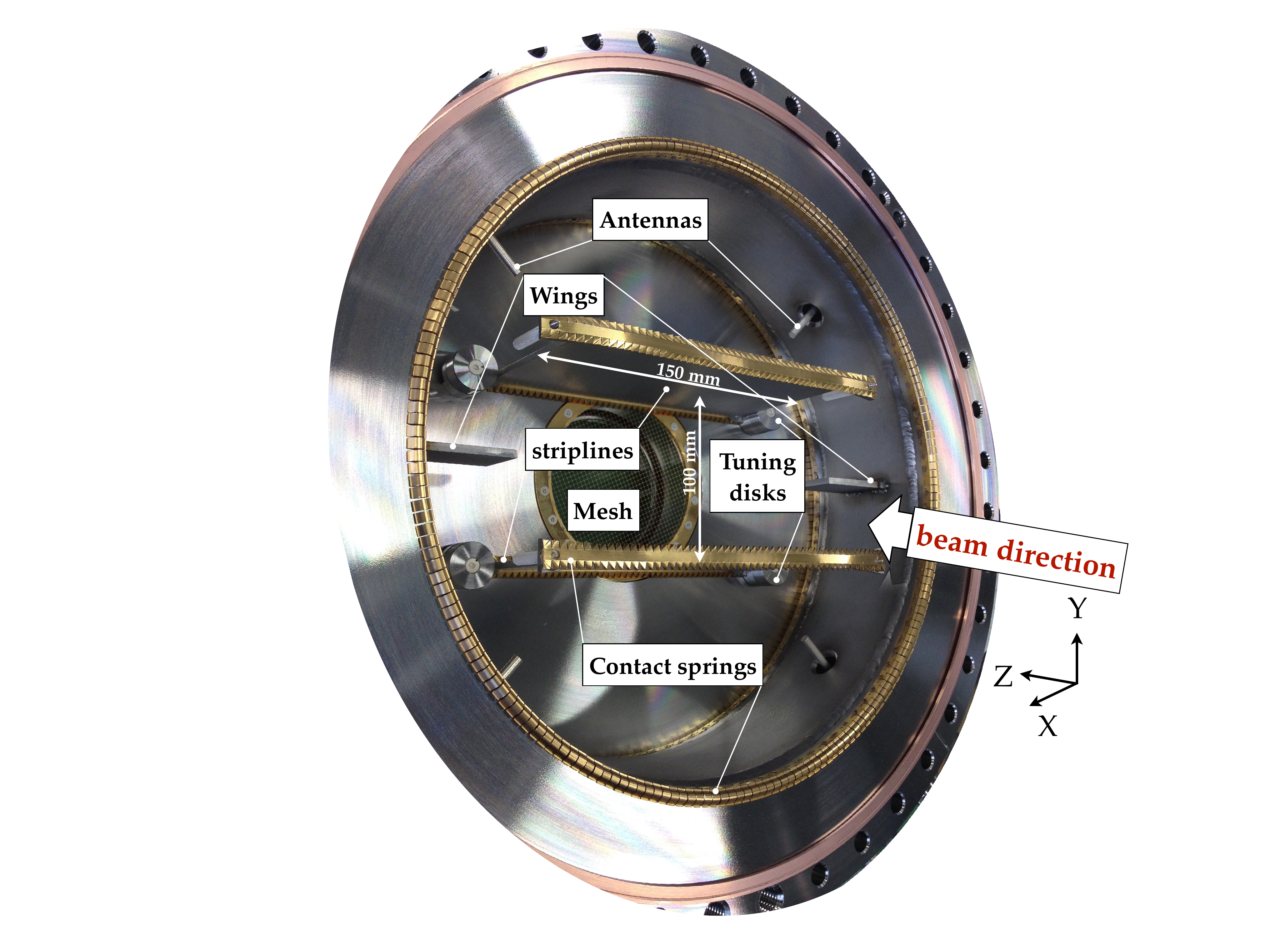}
\caption{Photograph of the cavity used to drive the hyperfine transition in hydrogen/antihydrogen. The central cavity body and the back closing flange are visible. The front closing flange is removed to reveal the inner structure of the cavity: the two strip lines resonators which are \unit[150]{mm} wide, \unit[105.5]{mm} long (in beam direction) and separated by a distance of \unit[100]{mm}, the wing structures and the tuning disks. The openings for the hydrogen/antihydrogen beam are closed-off by fine gold-plated mesh constraining the RF-field inside the cavity. The gold-plated contact springs ensure a good electrical contact between the closing flanges and the central body, as well as between the resonators and the flanges. The antennas assembled around the cavity for injection and pick-up of the RF-field  are connected via feedthroughs to the external circuit (see Fig.~\ref{f:Helm}). As shown in Fig.~\ref{fig:oddmode}, the magnetic field vector points in the x-direction.}
\label{fig:cavity}
\end{center}
\end{figure}

The cavity design has been detailed elsewhere \cite{Kroyer, Federmann_proceedings, Silkethesis}; in short it is a pillbox shaped strip line resonator (i.e composed of two parallel conducting plates) closed off by fine metallic meshes in the direction of the beam, allowing close to \unit[96]{\%} transparency for the incoming atoms while keeping the RF field from leaking out of the chamber, see Fig.~\ref{fig:cavity}.
The meshes were manufactured from a $\unit[100]{\mu m}$ thick sheet of stainless steel chemically etched to obtain the meshed structure of \unit[5]{mm} grid size, which was then gold-plated.\\
The cavity design was motivated by the requirement of a high RF-field homogeneity in the plane perpendicular to the beam axis.
The length of the conducting plates (measured along the beam, i.e. z direction) had to be an integer multiple of the desired resonance half-wavelength (it was chosen to be L$_{\rm cav}$=\unit[105.5]{mm} corresponding to half the wavelength of the zero-field hyperfine transition at \unit[1.42]{GHz}).
In contrast the distance between the plates could be chosen freely and was set to \unit[100]{mm}, matching the pipe diameter for a standard CF100 vacuum flange. 
Such a cavity consisting of a strip line inside a pillbox, can support two degenerate transverse electromagnetic (TEM) modes with similar resonant frequencies. Fig.~\ref{fig:oddmode} illustrates the magnetic field distribution of the two modes.
In order to retain only the desired odd mode, the even mode was suppressed in the frequency region of interest by addition of  ``wing''-structures (see Fig.~\ref{fig:cavity}) to selectively de-tune it.

\begin{figure}
\begin{center}
\includegraphics[trim = 3cm 1cm 3cm 1cm , width=0.8\textwidth]{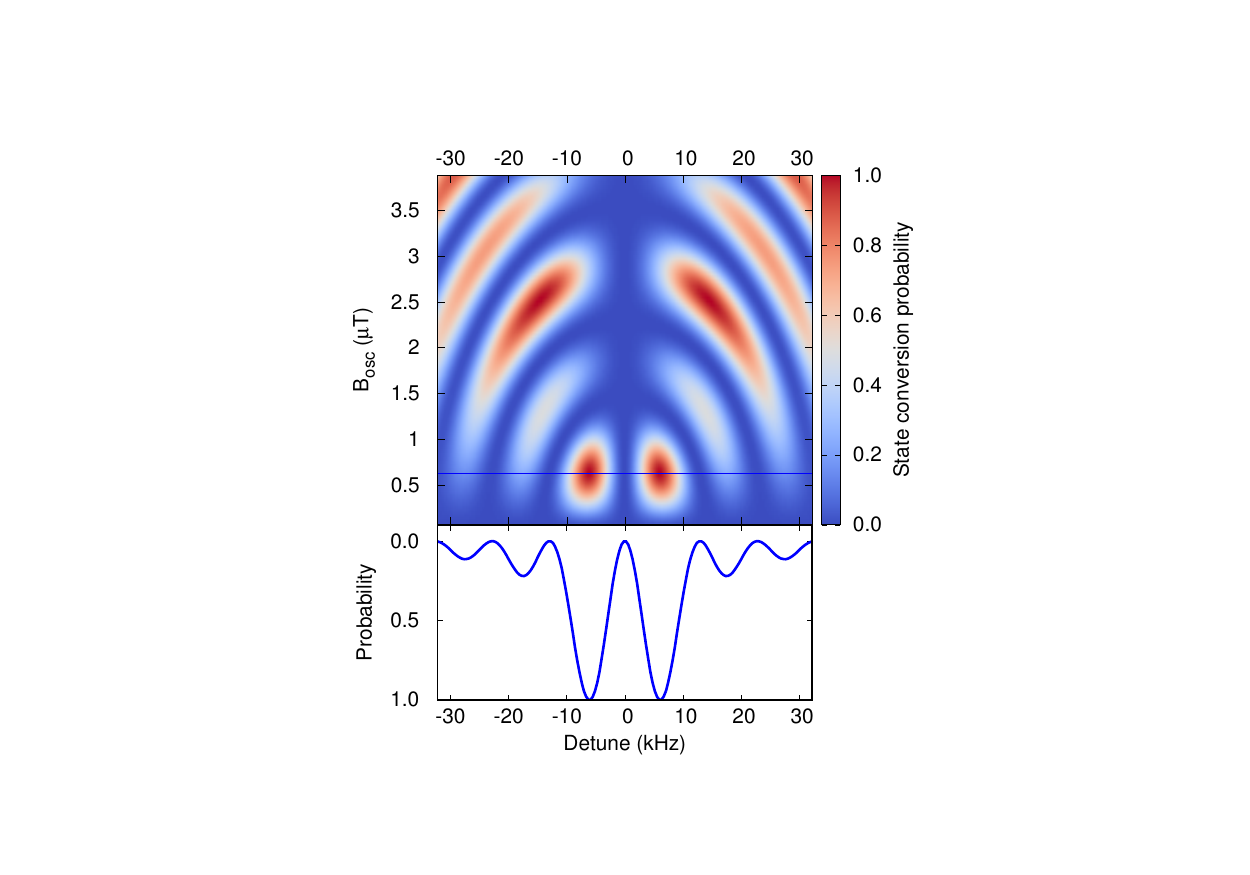}
\caption{Results of numerical solutions of the optical Bloch equations for a two-states system in a strip line cavity design. The state-conversion probability is given as a function of the de-tuning frequency and the amplitude of the oscillating magnetic field B$_{osc}$. A monoenergetic beam of \unit[1000]{ms$^{-1}$} and a cavity length of \unit[105.5]{mms} are assumed. The  horizontal line indicates the required driving strength to reach the first complete state conversion (``$\pi$-pulse''). The bottom plot is the  projection of the state-conversion probabilities at the line revealing the characteristic ``double-dip'' lineshape.}
\label{fig:double_dip}
\end{center}
\end{figure}

Simulations show that this geometry reached a field inhomogeneity in the plane perpendicular to the beam direction (x-y plane) of 2$\times(B_{\rm max}-B_{\rm min})/(B_{\rm max}+B_{\rm min})\sim  3\%$ \cite{Silkethesis}.
Along the beam (z direction), the field amplitude follows a sin($\textrm{z}\pi/\textrm{L}_{\rm cav}$) distribution: the field vanishes in the center of the cavity and has two maxima at the front and back walls of the resonator.
Thus the magnetic field points at opposite directions in the two half-volumes.
This field configuration leads to a ``double-dip'' structure in the resonance spectrum, as seen in Fig.~\ref{fig:double_dip}.   \\

\begin{figure}
\centering
\includegraphics[width=0.8\textwidth]{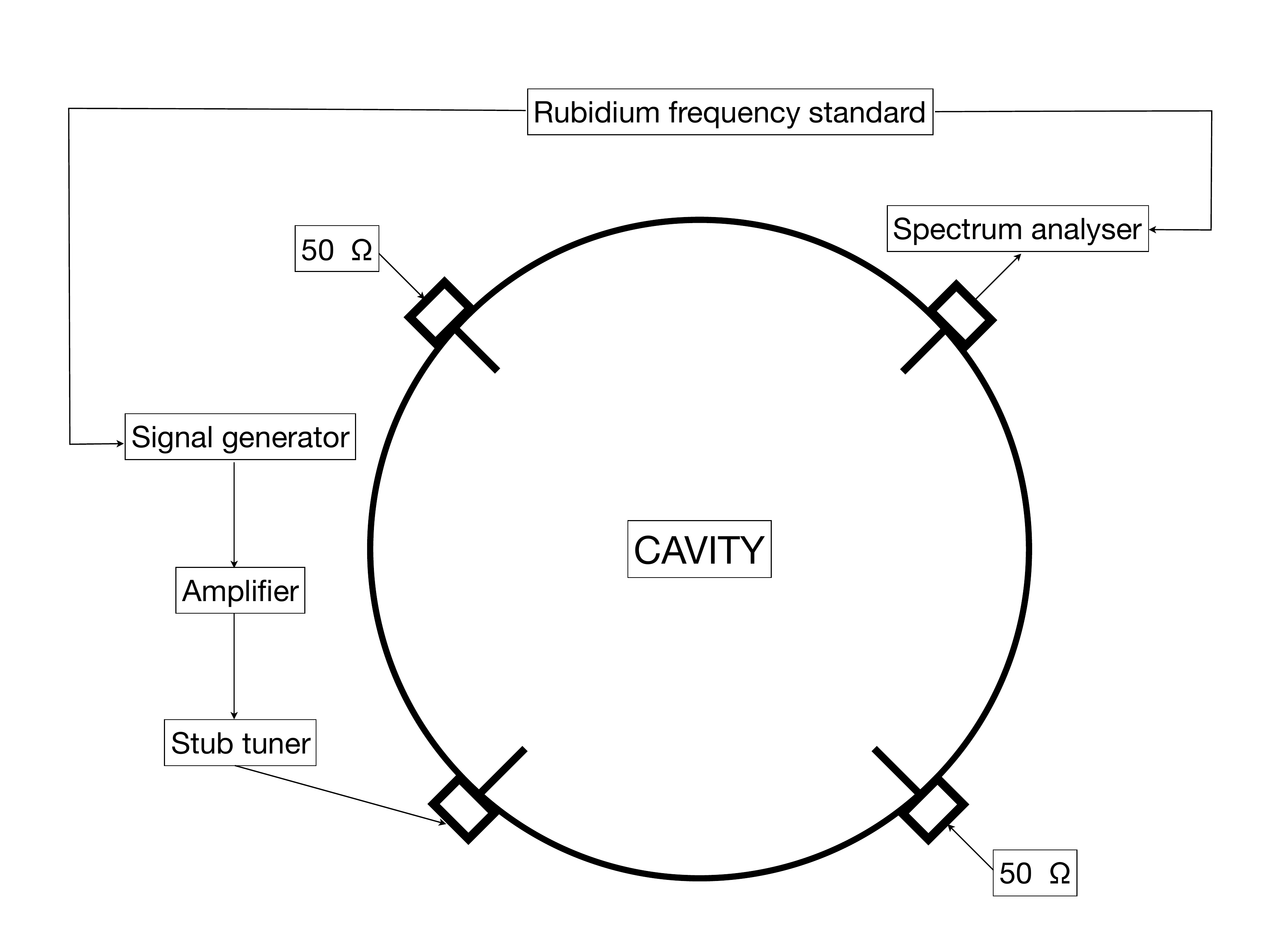}
\caption{Exciting scheme of the cavity. The RF field from a signal generator stabilized by a rubidium standard clock is amplified and injected via a stub tuner to one port of the cavity.
 The field is picked-up at another port and readout by a spectrum analyser also stabilized in frequency by the same rubidium clock. All other ports are terminated with \unit[50]{$\Omega$} connectors.}
\label{fig:cavity_excitation}
\end{figure}

\begin{figure}
\centering
\includegraphics[trim=0cm 6cm 0cm 5cm, clip=true, width=0.8\textwidth]{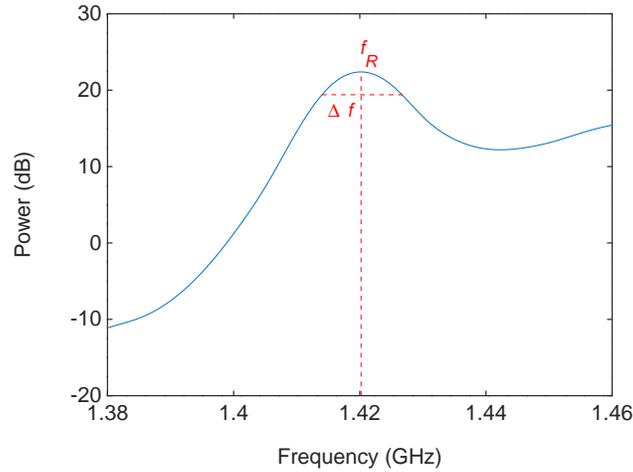}
\caption{Measured transmission scattering  parameter ($S_{21}$) between the readout and the input port revealing an isolated peak at the correct frequency and of the desired width.}
\label{fig:cavity_S21}
\end{figure}

The cavity consists of a stainless-steel central body closed off with copper seals by two modified CF400 flanges to match the high vacuum requirements of the setup. The cavity is equipped with four ports onto which UHV-compatible feedthroughs (PMB Alcen) are connecting antennas to the external apparatus.
The antennas lengths were adjusted to achieve an overcritical coupling necessary to reduce the quality factor of the cavity. This was required to retain a few MHz excitation range to measure the transitions at several external magnetic fields of the order of a few gauss (see \S\ref{s:Heml}).
In the original design two ports for feeding the microwaves in-phase and two ports for pick-up and analysis of the signal were envisioned. However, the interference pattern was found to be very sensitive to slightly different electrical lengths of the antennas. Therefore in a measurement run, one port was connected to a signal generator (Rohde \& Schwarz SML02), locked to a rubidium frequency standard (SRS FS725), via a \unit[42]{dB} amplifier (Mini Circuits ZHL-10W-2G(+)) and a stub-tuner (see discussion below). The opposing port was used to pick-up and monitor the signal using a spectrum analyzer (Agilent Technologies N9010A) connected to the same rubidium frequency standard and  the remaining two ports were terminated with \unit[50]{$\Omega$} connectors, as indicated in Fig.~\ref{fig:cavity_excitation}. \\
A comparison of the simulated transmission pattern and experimental measurements showed a detuning of the desired mode by \unit[26]{MHz}, necessitating the addition of tuning disks (see Fig.~\ref{fig:cavity}). It was however observed that the central frequency was more efficiently tuned by changing the distance between the meshes (influenced also by the pressure in the cavity, atmospheric pressure or vacuum) than by the tuning disks. The length of the cavity was therefore tuned by adding \unit[2]{mm} shims between the outer flange of the cavity and the closing meshes.\\
\indent Flexible gold-plated CuBe$_2$ contact springs (Fig.~\ref{fig:cavity}) ensure a good electrical connection between the two halves of the cavity and close the gap formed by the vacuum seal which would cause mode distortion.
Spot-welding of the springs onto the cavity structure was necessary to ensure a stability of the connections against movements and vibrations and a better reproducibility of the transmission pattern.
Nevertheless, external tunability and adjustment of the quality factor was required.
A coaxial double slug tuner (Microlab FXR SF-31N), providing a tunable impedance matching was thus inserted between the amplifier and the input port of the cavity. 
The typical microwave power injected into the cavity was of the order of \unit[0.3]{mW} to drive one half of a Rabi oscillation (``$\pi$-pulse'').\\
\indent Figure~\ref{fig:cavity_S21} shows the transmission scattering parameter $S_{21}$ (port 2 being the readout port and port 1 the input port) after impedance adjustments using the double slug tuner, where $f_R$ indicates the resonance frequency of the cavity and $\Delta f$ the width of the resonance at \unit[-3]{dB}.
The quality factor, Q=$\frac{f_R}{\Delta f}\sim$120 matches well the requirement of a \unit[4]{MHz} operational bandwidth to allow excitations within the range of 1420-\unit[1424]{MHz}.

\subsubsection{Helmholtz coils}\label{s:Heml}
\begin{figure}
\begin{center}
\includegraphics[width=0.8\textwidth]{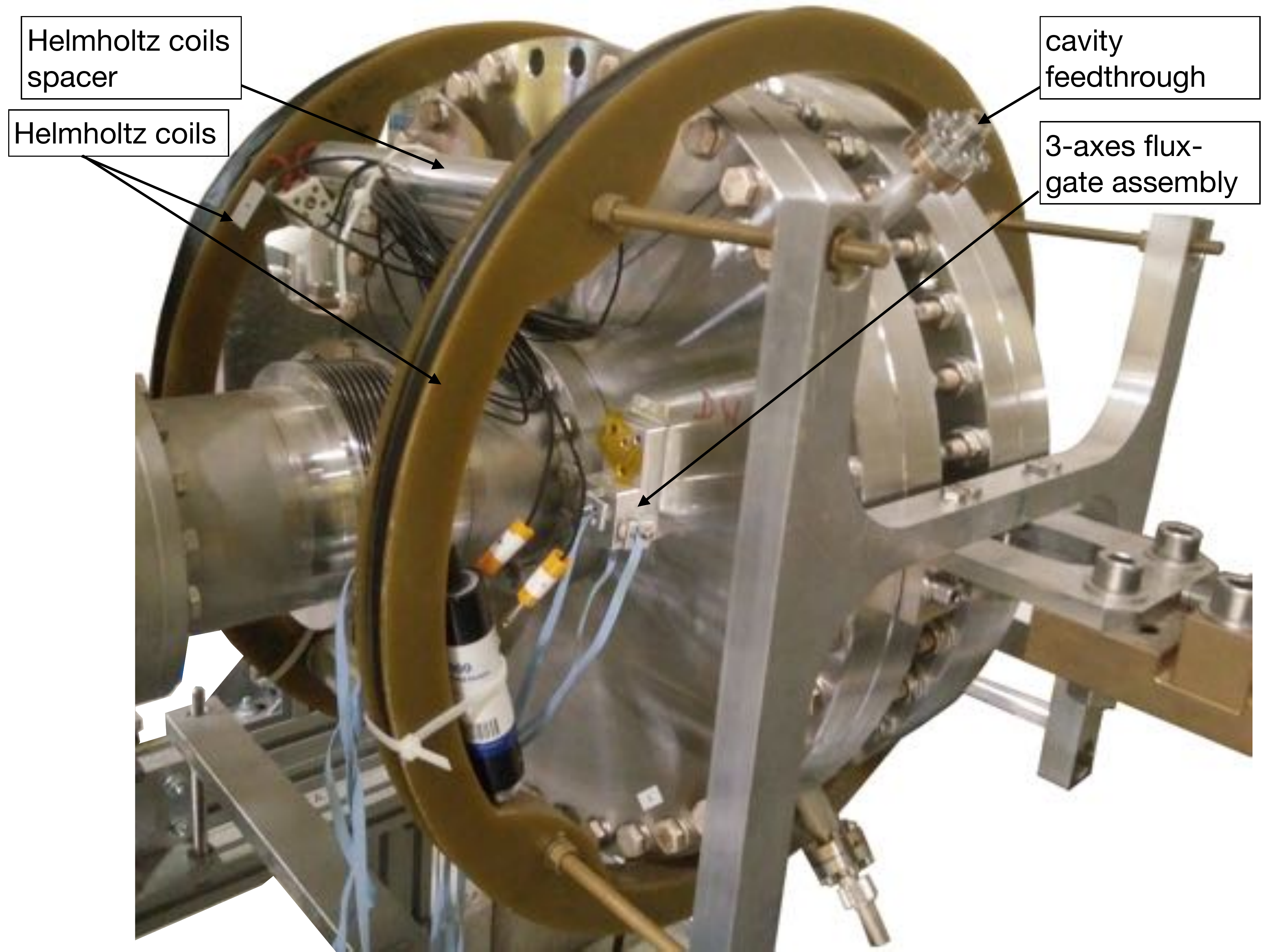} 
\caption{Photograph of the Helmholtz coils assembly mounted around the microwave cavity. One of the three-axes flux gate magnetometers (see \S\ref{s:magn_shield}) is visible on the flange of the cavity. The strip lines are aligned horizontally (not visible) leading to a parallel alignment of the oscillating and the static magnetic fields, as required to drive the $\sigma$-transition.}
\label{f:Helm}
\end{center}
\end{figure}

In the presence of a static magnetic field, the degeneracy of the hydrogen hyperfine triplet states is lifted as illustrated by the Breit-Rabi diagram in Fig.~\ref{fig:BreitRabi}. Several transitions are possible between the four hyperfine states which can be driven, depending on the relative orientation of the RF and static magnetic fields. In the setup reported here, a set of Helmholtz coils surrounded the spin flip cavity, producing a field perpendicular to the incoming particle beam and parallel to the oscillating RF field, see Fig.~\ref{f:Helm}. This configuration allows to drive the $\sigma_1$ transition (see Fig.~\ref{fig:BreitRabi}). The external magnetic field enters only in second order for this transition\footnote{$f_{\sigma_1}=\frac{\Delta E_0}{h}(1+\frac{1}{2}x^2)$ where $x=\frac{(g_J-g_I)\mu_B}{\Delta E_0}H$, $\Delta E_0$ is the transition energy in  zero-field, $\mu_B$ the Bohr magneton, $g_J$ , $g_I$ the electronic and nuclear g-values and $H$ the strength of the external magnetic field.}, making its determination less sensitive to field inhomogeneities. Therefore Helmholtz coils producing a field inhomogeneity ($\sigma_{|B|}/|B|$) better than 1\% in the volume of interest could be used.

The coils had radii of \unit[235]{mm} (at center of wiring, the innermost radius was \unit[220]{mm}) and 90 windings (10 rows, 9 windings/row). They were wound with a copper wire of \unit[1.6]{mm} in diameter onto a  support made of fiberglass loaded with epoxy. The coils were mounted on aluminum profiles fixed on the side of the cavity by threaded brass rod (see Fig.~\ref{f:Helm}), which, together with aluminum cylinders were used for accurate spacing of the coils. The distance between the coils was optimized in the presence of a magnetic shielding, described in \S\ref{s:magn_shield}, to produce the most homogeneous field in the region of interest. The optimal distance was found to be \unit[214]{mm} which is slightly smaller than the design radius of the coils due to the presence of the shielding.\\
The current to the coils, connected in series, was delivered by a stable power supply (Heinzinger PTNhp) and monitored with a digital multimeter (Keithley 2001). The current was set between \unit[$\pm$1]{A}, corresponding to a maximum magnetic field of \unit[459]{$\mu$T} and a change in the $\sigma_1$ transition frequency of a few tens of kHz (see Fig.\ref{fig:HFS_extrapolation}).   

\subsubsection{Magnetic shielding}
\label{s:magn_shield}
To minimize the influence of external stray magnetic fields, such as the Earth's or the downstream sextupole magnet's, in the interaction region, a cuboid two-layer magnetic shielding made of mu-metal was built and assembled around the cavity and Helmholtz coils.
The outer dimensions (width $\times$ length $\times$ height) of the inner and outer layer were $531\times531\times606~\unit{mm^3}$ and $561\times561\times636~\unit{mm^3}$ with a thickness of \unit[1]{mm} and \unit[2]{mm}, respectively. 
Two three-axes flux gate magnetometers (Bartington Mag-03IE1000 read out with Bartington Decaport) were placed on each side of the cavity to monitor the magnetic field and any field fluctuations inside the shielding (Fig.\ref{f:Helm}).
The shielding factor, the ratio of inside- to outside-shielding magnetic field strengths measured, of the configuration depends on the orientation of the external field, as the layers have openings larger than \unit[100]{mm} in diameter in the axial direction, where the beam pipe enters and exits.
A COMSOL simulation, assuming a mu-metal relative permeability of $\mu_r > 20,000$, gave shielding factors of typically 150 for static axial fields and 800 for static radial fields.
To experimentally determine the shielding factor one additional flux gate magnetometer was placed outside the shielding to monitor the inside and outside fields in parallel.
This was done when the apparatus was installed at the ASACUSA antihydrogen setup in the hall of the AD. However, only the inner shielding layer was present for this measurement.
The changes of the internal and external stray fields related to the cycle of the AD ($\approx$\unit[100]{s}) could be monitored. A radial shielding factor of $\approx$42 was measured. Similar measurements were performed at the hydrogen setup during a ramp-up of the superconducting sextupole and a radial shielding factor of $\approx$54 and $\approx$1470 was measured for a single layer and two layers of shielding respectively. Axial shielding factors are highly dependent on the configuration of the external field and are thus difficult to compare experimentally. The higher measured radial shielding factor than simulated can be explained by the conservative $\mu_r$ chosen for mu-metal in the simulation.      
\subsubsection{Superconducting Sextupole}
\label{sec:sc6pole}
The spin-state analysis after the interaction is performed through magnetic sextupole fields, similar to the initial beam polarizer. 
In the radial plane this field configuration can be parametrized by
\begin{equation}
\vec{B}_r=\frac{B_0}{r^2_0} \cdot 
\left(
\begin{array}{c}
x^2 - y^2 \\
- 2 x y 
\end{array} \right)
\end{equation}
where $B_0$ denotes the field strength at radius $r_0$.
The absolute value of the magnetic field $|B|$ is proportional to $r^2$, the force on the magnetic moment of a particle is proportional to the gradient $\vec{\nabla}|\vec{B}|$.
Therefore sextupole magnets produce forces which point radially and scale linearly with the distance from the beam axis in the regime of field-independent moments.
As a consequence the particles follow harmonic trajectories in the radial plane (see Fig.~\ref{fig:sextupole_doublet}).
Such a field is characterized by the constant ratio of the gradient to the radius
\begin{equation}
G' = \frac{1}{r}\frac{\partial|\vec{B}|}{\partial r} = 2 \frac{B_0}{r^2_0} .
\end{equation}
In contrast to normal Stern-Gerlach magnets with constant gradients a sextupole configuration enables two-dimensional focusing (defocusing) of a beam of lfs (hfs), similar to optical lenses. 

The combination of the quadratic radius dependence and the requirement of a large acceptance for antihydrogen (aperture of \unit[100]{mm}) translates into strong magnetic fields at the poles, hence supplied by superconducting coils designed and constructed by Tesla Engineering Ltd.
The coils were placed in a vacuum isolated tank filled with liquid helium.
The heat load on the inner vessel is absorbed by a cryocooler which prevents the boil off of the helium and enables continuous operation.
The warm bore ($\sim\unit[90]{K}$) serves directly as the vacuum pipe of the beam experiment and can be connected to the adjacent components via standard CF100 flanges.
A power supply (SMS550C from Cryogenic Limited) provides currents up to \unit[400]{A} to the multi-layer bent racetrack coil geometry.
The design magnetic field of $B_0$=\unit[2.42]{T} is produced at a radius of $r_0$=\unit[50]{mm} at the maximum current.
The gradient-constant is $G'$=\unit[1936]{Tm$^{-2}$}.
The focusing strength of a sextupole magnet of effective length $L$ is given by the axial integral over the gradient constant $\int G'(z) dz$.
With $L=\unit[220]{mm}$ the expected integrated gradient amounts to $G'dL$=\unit[436]{Tm$^{-1}$} at \unit[400]{A}. A measurement at room temperature and at \unit[0.8]{A} yielded, after scaling to \unit[400]{A}, $G'dL$=\unit[515]{Tm$^{-1}$}, which is even higher than the design value.

\subsection{Hydrogen Detection}
\label{s:H_detection}
\begin{figure}
\begin{center}
\includegraphics[trim=5cm 0cm 3cm 0cm, clip=true,width=0.8\textwidth]{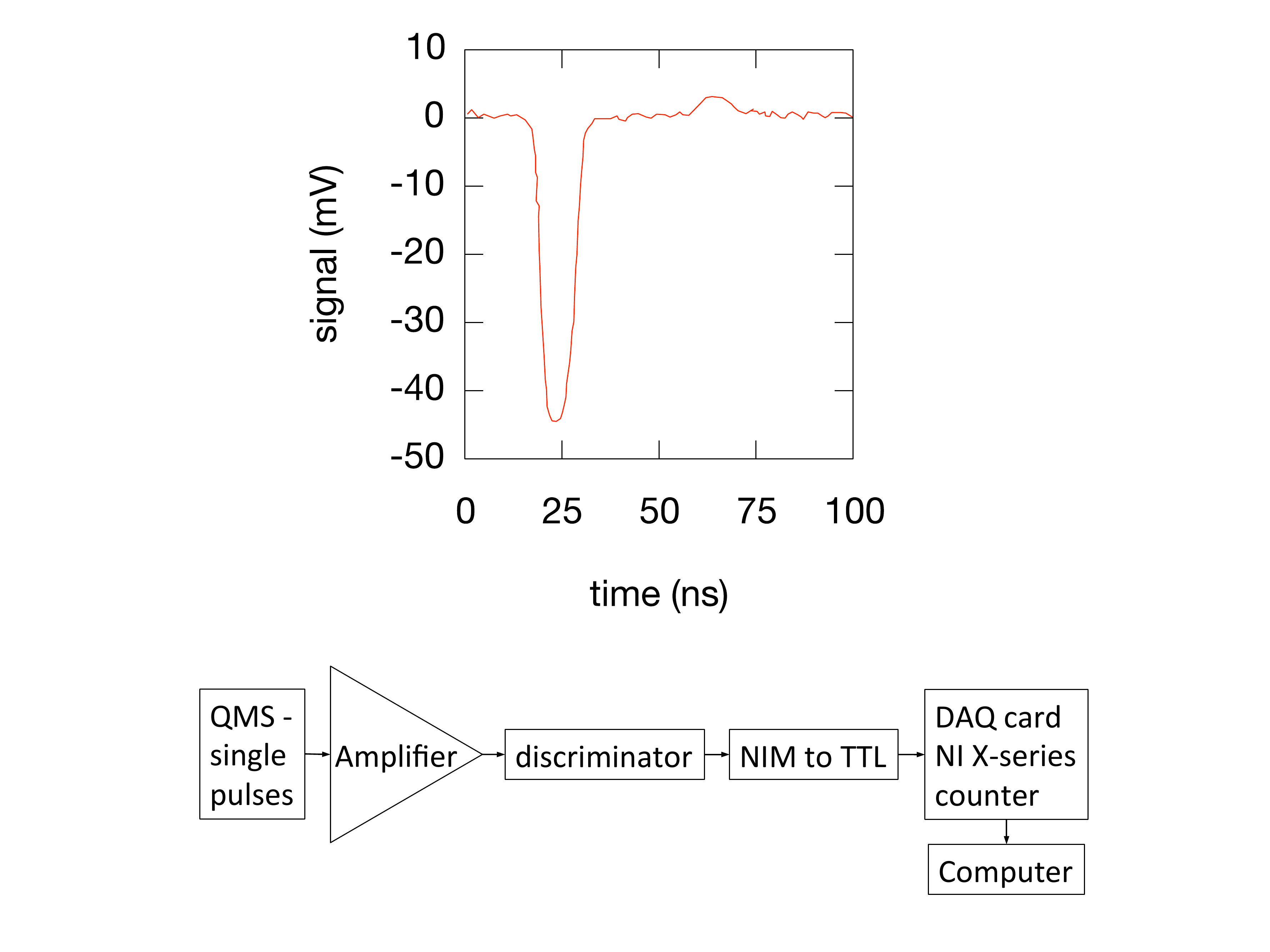}
\caption{Top: typical single ion pulse from the SCEM measured with an oscilloscope. Bottom: schematic of the readout chain.}
\label{fig:DAQ_single_counts}
\end{center}
\end{figure}

A quadrupole mass spectrometer (QMS) (MKS Microvision 2 100 D in crossed beam configuration) was placed at the end of the beamline and tuned to identify particles which atomic mass is one to detect atomic hydrogen. The overall efficiency of the QMS for detection of hydrogen was around $10^{-8}$.
The QMS was mounted onto translational stages which could move the detector in a plane perpendicular to the incoming beam for investigating beam profiles and signal optimization. The typical area scanned was about \unit[112]{mm$^2$} ($\sim\pm\unit[4]{mm}$ in the horizontal direction and $\pm\unit[7]{mm}$ in the vertical direction) and the typical step sizes were of the order of a millimeter, which is much larger than the minimum step sizes the stage could perform.
The on-board single channel electron multiplier (SCEM), powered with an external high voltage, amplified the signal which was fed into another external amplifier (LeCroy - LRS 333). The total signal amplification was of the order of \unit[20]{dB}. The signal was then discriminated, converted to TTL and read out by a data acquisition card (NI ADC module: PCIe-6361). Figure \ref{fig:DAQ_single_counts} shows a typical pulse after amplification as well as the signal readout scheme.\\
\indent High molecular hydrogen background decreases the atomic beam detection efficiency. Both active and passive background suppression methods were therefore used.
A non-evaporable getter pump (NEG), in combination with two-stage turbo molecular pumping, enabled a vacuum level better than \unit[5$\times10^{-8}$]{Pa} in the QMS chamber. Additionally, digital data processing using lock-in amplification techniques enabled further discrimination against background. Information on the phase of  the tuning fork chopper (see \S\ref{s:chopper}) was obtained by feeding the monitor signal of the chopper driver to a sampling voltage input of the NI ADC module. A software algorithm separated the modulated content of the count rate from the constant background rate \cite{WOL13}.\\

\section{Characterization measurements}
\label{s:characterization}
\subsection{Time-of-flight and velocity measurements}\label{s:TOF}
\begin{figure}
\begin{center}
\includegraphics[width=0.8\textwidth]{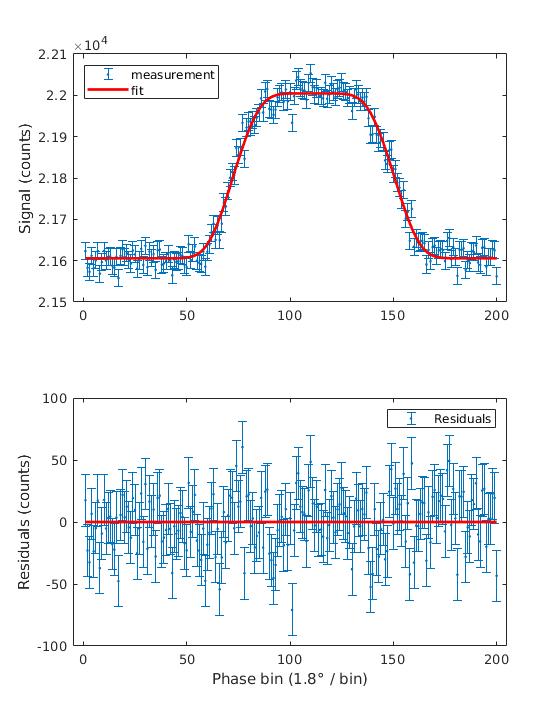}
\caption{Top: count rate at the QMS vs phase of the chopper signal. The fit function (red solid line) is a truncated positive half-wave of a sine convoluted with a gaussian. Bottom: residuals of fit.}
\label{fig:chopper_phase}
\end{center}
\end{figure}

The time-of-flight $t$ of the hydrogen atoms was determined by relating the time-dependent beam count rate at the QMS detector and the time-dependent signal of the alignment laser on a photo-diode (as a proxy for the chopper opening) to the phase of the chopper reference signal recorded by the software lock-in amplifier. From those measurements the mean velocity $v$ of the beam was extracted in the following way: $v$ is the time-of-flight of the hydrogen atoms over the distance $L$ between the chopper and the QMS detector, which is evaluated using the difference between the phase of a laser $\Phi_L$ shining through the entire apparatus (the time-of-flight being negligible) and the phase of the beam signal $\Phi$, over the chopper reference signal with frequency $f_\textrm{ch}$: 
\begin{equation}\label{eq:velocity}
v = \frac{L}{t} = \frac{2\pi\times L \times f_{\rm ch}}{\Phi-\Phi_L}
\end{equation}
where $\Phi_L$  and $\Phi$ are expressed in radian and $f_{\rm ch}$ is extracted from the lock-in amplifier software. The phase $\Phi_L$ is determined from the laser signal, read out by a photodiode behind the QMS detector, while $\Phi$ is evaluated using the beam count rate measured at the QMS as a function of the phase of the chopper (Fig.\ref{fig:chopper_phase}). The count rate has two components: a constant background and a sinusoid-shaped modulated beam signal. 
Several fits to extract $\Phi$ were attempted. The one giving the best results in terms of goodness-of-fit was a truncated positive sine half-wave convoluted with a gaussian. The truncation comes from the size of the beam with respect to the chopper opening, while the convolution with the gaussian takes into account the velocity distribution of the beam. Figure~\ref{fig:chopper_phase} shows a typical histogram and the residuals of such a fit.\\
\indent The knowledge of $\Phi$, $\Phi_L$, $f_{\rm ch}$ and $L$ yielded, using (\ref{eq:velocity}), the velocities plotted for example in Fig.~\ref{fig:sextupole_velocity_selection_sim_data} where the error bars take into account the errors on the laser phase, on the beam phase (being the dominant ones), on the length $L$ and on the chopper frequency $f_{\rm ch}$.


\subsection{Focusing by the superconducting sextupole}
\begin{figure}
\centering
\includegraphics[width=0.8\textwidth, trim={2cm 0.2cm 2cm 0.2cm},clip]{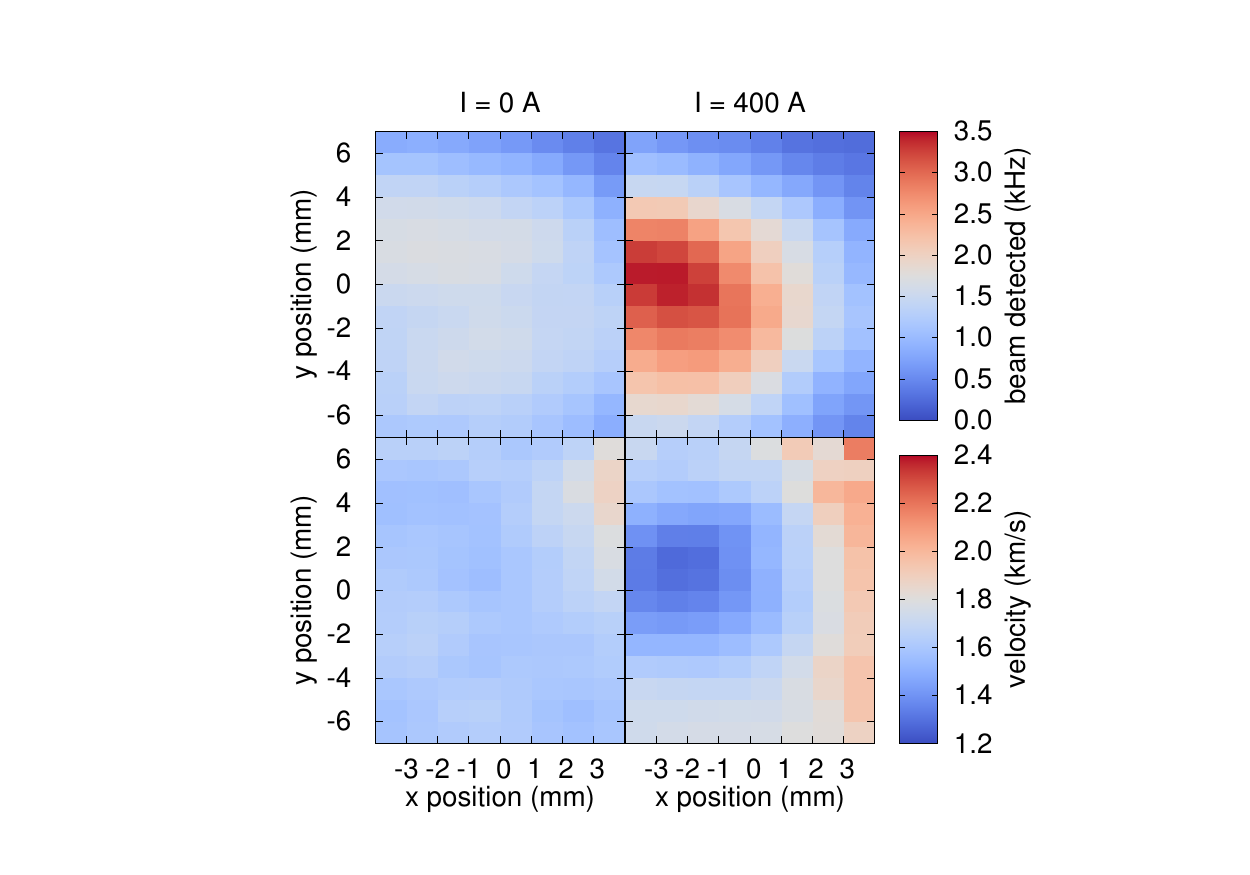}
\caption{Focusing effect of the analyzing superconducting sextupole; top (bottom) panels compare the beam count rate (average beam velocity) when the superconducting sextupole is turned off (left panels) or energized with the maximum current of \unit[400]{A} (right panels). The data were collected in the straight-source configuration.}
\label{fig:sextupole_focussing}
\end{figure}

\begin{figure}
\centering
\includegraphics[width=0.8\textwidth, trim={2cm 0.2cm 2cm 0.2cm},clip]{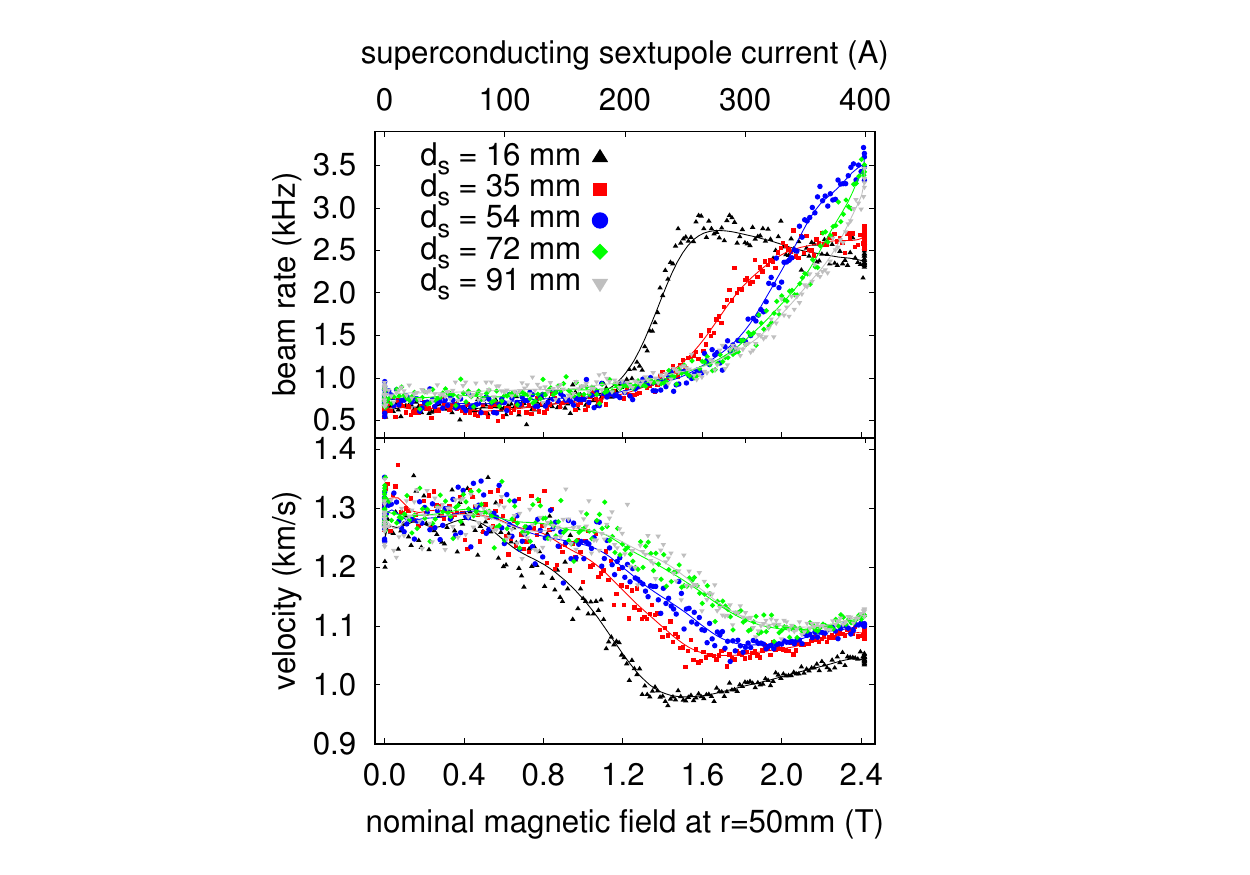}
\caption{Measured count rates (top) and average beam velocities (bottom) while ramping the superconducting sextupole magnet from zero up to the maximum current of \unit[400]{A} (the second x-axis gives the corresponding magnetic field strength at a radius of \unit[50]{mm}). The trends are compared for 5 different settings of the distance $d_s$ between the sextupole doublet, which pre-selects the beam velocity. The solid lines indicate the running average. The data were collected in the \unit[90]{$\degree$} source configuration.}
\label{fig:sextupole_focus}
\end{figure}

The effect of the superconducting sextupole can be characterized using the 2D-scanning stages (\S\ref{s:H_detection}) and a measurement of the time-of-flight of the atoms arriving at the QMS. Figure~\ref{fig:sextupole_focussing} shows  two-dimensional beam profiles indicating the position dependence of either the count rate or the velocity of the hydrogen beam recorded at the QMS for sextupole current off and for \unit[400]{A} at the maximum $d_\textrm{s}$ (highest initial velocity selection). It clearly illustrates the focusing of the low velocity components when the magnet is energized.  
Figure~\ref{fig:sextupole_focus} also illustrates this behavior at different initial velocities (5 different $d_\textrm{s}$ spacings). At lower velocities the sextupole current needed to focus a high portion of the low velocity component is lower. 

\subsection{Measurement of the hyperfine transition}
\begin{figure}
\centering
\includegraphics[clip, width=0.4\textwidth]{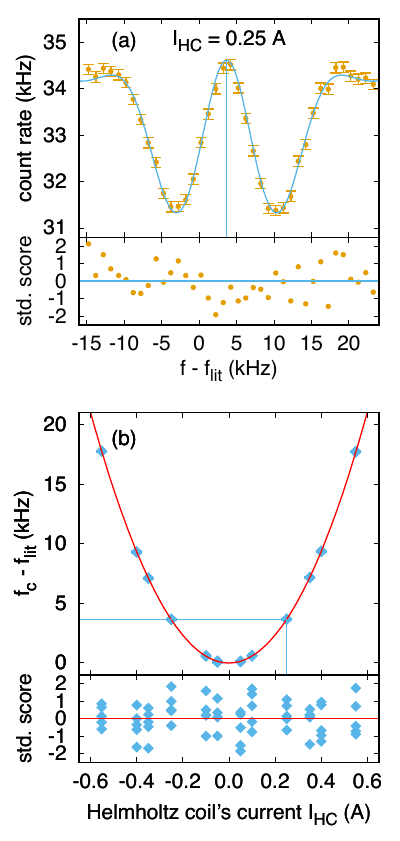}
\caption{Determination of the ground-state hyperfine splitting of hydrogen; (a) the lineshape was measured at different external magnetic fields/Helmholtz coils currents (here an example at \unit[250]{mA}) and provided the frequency of the $\sigma_1$ transition at those fields indicated by the blue vertical line; (b) the extrapolations of those values to zero magnetic field yielded the zero-field transition with a relative precision of $3\times10^{-9}$.}
\label{fig:HFS_extrapolation}
\end{figure}

In \cite{ourNatCommuns} we reported the measurement, using this apparatus, of the ground-state hyperfine splitting of hydrogen through the $\sigma_1$ transition with a relative precision of $3\times10^{-9}$, which constitutes more than an order of magnitude improvement over previous determinations in a beam \cite{Kusch1,Kusch2}. The ground-state value was obtained by extrapolating several measurements taken at different external magnetic fields (within the few gauss range) to zero-field. Figure~\ref{fig:HFS_extrapolation} shows such an extrapolation.
The investigation of systematic shifts included, among others, the size of the beam at the cavity entrance (see \S\ref{s:spectro_app}), the effect of the shielding factor, the temperature of the source, and the orientation of the PTFE tubing. The only potential systematic shift that could be identified stems from the frequency standard and was on the \unit[1]{ppb} level.
All results were in agreement, within their uncertainties, with the more precise literature value ($f_{\rm lit}=\unit[1,420,405,751.768 (2)]{Hz}$ \cite{Hellwig,Karshenboim}). Therefore systematic shifts stemming from the spectroscopy method or apparatus can be excluded at the few ppb level for the planned measurements on antihydrogen with an initial precision goal of \unit[1]{ppm}.

\section{Summary and Conclusions}
We have described  ASACUSA's hydrogen apparatus to commission and characterize the spectroscopy setup envisioned for the measurement of the ground-state hyperfine splitting of antihydrogen. The setup, designed for high acceptance of the scarce antihydrogen atoms, is capable of reaching the specified performance to selectively drive the $\sigma_1$ transition. A measurement of the ground-state hyperfine splitting of hydrogen was performed \cite{ourNatCommuns} with an order of magnitude improved precision over previous determination in hydrogen beams. The prospects for antihydrogen spectroscopy depend on the quality of the antihydrogen beam and in particular on the average velocity, polarization and quantum states of the atoms exiting the formation region toward the apparatus. An evaluation of the needed amount of antihydrogen atoms given particular values of those parameters to reach a ppm precision on antihydrogen has been provided in \cite{ourNatCommuns}.\\ 
\indent An upgrade (improved coils configuration and shielding) of the apparatus described in this manuscript has been performed to allow the simultaneous determination of the $\sigma_1$ and $\pi_1$ transitions and will be the subject of a future publication. This is of particular interest for the antihydrogen measurement since a single measurement of the two transitions at a given field is sufficient to determine the ground-state hyperfine splitting.

\section*{Acknowledgments}
This work was funded by the European Research Council under European Union's Seventh Framework Programme (FP7/2007- 2013)/ERC Grant agreement (291242) and the Austrian Ministry of Science and Research, Austrian Science Fund (FWF) DK PI (W 1252), and supported by the CERN fellowship and summer student programmes as well as the DAAD RISE programme.\\
The authors wish to thank P.~Caradonna for his initial contributions to the experimental setup, as well as B.~Juhasz for proposing the idea of velocity selection via a sextupole doublet, and B.~W\"unschek for performing the initial simulations. The authors are also indebted to S.~Federmann, F.~Caspers, T.~Kroyer and several members of the CERN BE department (previously in the AB department) and TE-MPE-EM groups, who heavily contributed to the design and manufacturing of the cavity. We are grateful to P.~A.~Thonet and D.~Tommasini from the CERN TE-MSC-MNC group for their help in the design and manufacturing of the sextupole doublet and the Helmholz coils, respectively.
We acknowledge technical support by the CERN Cryolab and Instrumentation group TE-CRG-CI, especially L.~Dufay-Chanat, T.~Koettig, and T.~Winkler.
The hydrogen source was gracefully provided by the group of H.~Knudsen. We thank H.-P.~E.~Kristiansen for providing initial support for the assembly and operation.
We also wish to thank C.~Jepsen, J.~Hansen, M.~Wolf, M.~Heil, F.~Pitters, Ch.~Klaushofer and S.~Friedreich who contributed to the characterization of the apparatus.



\section*{Author's contributions}

C.A, H.B., C.E, M.F., B.K., M.L., V.M., C.M., V.M., O.M., Y.M., Y.N., C.S., M.C.S, L.V., E.W., Y.Y and J.Z. are involved in the ASACUSA-CUSP antihydrogen program. S.A.C., M.D., C.M., O.M., M.C.S, E.W, M.W. and J.Z. were involved in the hydrogen experiment of which apparatus is described in this manuscript.  All authors contributed to critical discussions regarding the published content.
The spectroscopy apparatus was built and operated by M.D, C.M, O.M., M.C.S and J.Z. The measurements described in this manuscript were taken by M.D, C.M, M.C.S and in parts by M.W.. B.K. and C.S. performed simulations. E.W. proposed the experiment. 
C.M. wrote the manuscript which was critically reviewed by all authors.





\bibliographystyle{elsarticle-num} 
\bibliography{literature}

\begin{thebibliography}{10}
\expandafter\ifx\csname url\endcsname\relax
  \def\url#1{\texttt{#1}}\fi
\expandafter\ifx\csname urlprefix\endcsname\relax\def\urlprefix{URL }\fi
\expandafter\ifx\csname href\endcsname\relax
  \def\href#1#2{#2} \def\path#1{#1}\fi

\bibitem{ALP182}
M.~Ahmadi, B.~X.~R. Alves, C.~J. Baker, W.~Bertsche, A.~Capra, C.~Carruth,
  C.~L. Cesar, M.~Charlton, S.~Cohen, R.~Collister, S.~Eriksson, A.~Evans,
  N.~Evetts, J.~Fajans, T.~Friesen, M.~C. Fujiwara, D.~R. Gill, J.~S. Hangst,
  W.~N. Hardy, M.~E. Hayden, C.~A. Isaac, M.~A. Johnson, J.~M. Jones, S.~A.
  Jones, S.~Jonsell, A.~Khramov, P.~Knapp, L.~Kurchaninov, N.~Madsen,
  D.~Maxwell, J.~T.~K. McKenna, S.~Menary, T.~Momose, J.~J. Munich,
  K.~Olchanski, A.~Olin, P.~Pusa, C.~{\O}. Rasmussen, F.~Robicheaux, R.~L.
  Sacramento, M.~Sameed, E.~Sarid, D.~M. Silveira, G.~Stutter, C.~So, T.~D.
  Tharp, R.~I. Thompson, D.~P. van~der Werf, J.~S. Wurtele, Characterization of
  the {1S$–$2S} transition in antihydrogen, Nature 557 (2018) 71--75.

\bibitem{ALPHA2017_2}
M.~Ahmadi, B.~X.~R. Alves, C.~J. Baker, W.~Bertsche, E.~Butler, A.~Capra,
  C.~Carruth, C.~L. Cesar, M.~Charlton, S.~Cohen, R.~Collister, S.~Eriksson,
  A.~Evans, N.~Evetts, J.~Fajans, T.~Friesen, M.~C. Fujiwara, D.~R. Gill,
  A.~Gutierrez, J.~S. Hangst, W.~N. Hardy, M.~E. Hayden, C.~A. Isaac,
  A.~Ishida, M.~A. Johnson, S.~A. Jones, S.~Jonsell, L.~Kurchaninov, N.~Madsen,
  M.~Mathers, D.~Maxwell, J.~T.~K. McKenna, S.~Menary, J.~M. Michan, T.~Momose,
  J.~J. Munich, P.~Nolan, K.~Olchanski, A.~Olin, P.~Pusa, C.~{\O}. Rasmussen,
  F.~Robicheaux, R.~L. Sacramento, M.~Sameed, E.~Sarid, D.~M. Silveira,
  S.~Stracka, G.~Stutter, C.~So, T.~D. Tharp, J.~E. Thompson, R.~I. Thompson,
  D.~P. van~der Werf, J.~S. Wurtele,
  \href{http://dx.doi.org/10.1038/nature23446}{Observation of the hyperfine
  spectrum of antihydrogen}, Nature 548 (2017) 66 EP --.
\newline\urlprefix\url{http://dx.doi.org/10.1038/nature23446}

\bibitem{ALPHA2018_Lyman}
M.~Ahmadi, B.~X.~R. Alves, C.~J. Baker, W.~Bertsche, A.~Capra, C.~Carruth,
  C.~L. Cesar, M.~Charlton, S.~Cohen, R.~Collister, S.~Eriksson, A.~Evans,
  N.~Evetts, J.~Fajans, T.~Friesen, M.~C. Fujiwara, D.~R. Gill, J.~S. Hangst,
  W.~N. Hardy, M.~E. Hayden, E.~D. Hunter, C.~A. Isaac, M.~A. Johnson, J.~M.
  Jones, S.~A. Jones, S.~Jonsell, A.~Khramov, P.~Knapp, L.~Kurchaninov,
  N.~Madsen, D.~Maxwell, J.~T.~K. McKenna, S.~Menary, J.~Michan, T.~Momose,
  J.~J. Munich, K.~Olchanski, A.~Olin, P.~Pusa, C.~{\O}. Rasmussen,
  F.~Robicheaux, R.~L. Sacramento, M.~Sameed, E.~Sarid, D.~M. Silveira, D.~M.
  Starko, G.~Stutter, C.~So, T.~D. Tharp, R.~I. Thompson, D.~P. van~der Werf,
  J.~S. Wurtele, Observation of the 1s–2p lyman-$\alpha$ transition in
  antihydrogen, Nature 561 (2018) 211--215.

\bibitem{Hellwig}
H.~Hellwig, R.~F. Vessot, M.~W. Levine, P.~W. Zitzewitz, D.~W. Allan, D.~J.
  Glaze, Measurement of the unperturbed hydrogen hyperfine transition
  frequency, Instrumentation and Measurement, IEEE Transactions on 19~(4)
  (1970) 200--209.

\bibitem{Karshenboim}
S.~G. Karshenboim, Some possibilities for laboratory searches for variations of
  fundamental constants, Canadian Journal of Physics 78~(7) (2000) 639--678.

\bibitem{Essen1}
L.~Essen, R.~W. Donaldson, M.~J. Bangham, E.~G. Hope, Nature 229 (1971) 110.

\bibitem{Essen2}
L.~Essen, R.~Donaldson, E.~Hope, M.~Bangham, Hydrogen maser work at the
  national physical laboratory, Metrologia 9~(3) (1973) 128.

\bibitem{Ramsey1}
N.~Ramsey, Atomic hydrogen hyperfine structure experiments, in: T.~Kinoshita
  (Ed.), Quantum Electrodynamics, World Scientific, Singapore, 1990, pp.
  673--695.

\bibitem{Ramsey2}
N.~F. Ramsey,
  \href{http://link.aps.org/doi/10.1103/RevModPhys.62.541}{Experiments with
  separated oscillatory fields and hydrogen masers}, Rev. Mod. Phys. 62 (1990)
  541--552.
\newblock \href {http://dx.doi.org/10.1103/RevModPhys.62.541}
  {\path{doi:10.1103/RevModPhys.62.541}}.
\newline\urlprefix\url{http://link.aps.org/doi/10.1103/RevModPhys.62.541}

\bibitem{Rabi1}
I.~I. Rabi, J.~R. Zacharias, S.~Millman, P.~Kusch, {A New Method of Measuring
  Nuclear Magnetic Moment$^{ }$}, Physical Review 53 (1938) 318.

\bibitem{Rabi2}
I.~I. Rabi, S.~Millman, P.~Kusch, J.~R. Zacharias, {The Molecular Beam
  Resonance Method for Measuring Nuclear Magnetic Moments. The Magnetic Moments
  of $_ {3} $Li$^{6}$, $_ {3}$ Li$^{7}$ and $_ {9}$ F$^{19}$}, Physical Review
  55~(6) (1939) 526.

\bibitem{Clemens_diss}
C.~Sauerzopf,
  \href{http://repositum.tuwien.ac.at/obvutwhs/content/titleinfo/1327619}{The
  {ASACUSA} antihydrogen detector: Development and data analysis}, Ph.D.
  thesis, Technische Universit{\"a}t Wien (2016).
\newline\urlprefix\url{http://repositum.tuwien.ac.at/obvutwhs/content/titleinfo/1327619}

\bibitem{Kusch1}
A.~G. Prodell, P.~Kusch, The hyperfine structure of hydrogen and deuterium,
  Physical Review 88~(2) (1952) 184.

\bibitem{Kusch2}
P.~Kusch,
  \href{http://link.aps.org/doi/10.1103/PhysRev.100.1188}{Redetermination of
  the hyperfine splittings of hydrogen and deuterium in the ground state},
  Phys. Rev. 100 (1955) 1188--1190.
\newblock \href {http://dx.doi.org/10.1103/PhysRev.100.1188}
  {\path{doi:10.1103/PhysRev.100.1188}}.
\newline\urlprefix\url{http://link.aps.org/doi/10.1103/PhysRev.100.1188}

\bibitem{ourNatCommuns}
M.~Diermaier, C.~B. Jepsen, B.~Kolbinger, C.~Malbrunot, O.~Massiczek,
  C.~Sauerzopf, M.~C. Simon, J.~Zmeskal, E.~Widmann,
  \href{http://dx.doi.org/10.1038/ncomms15749}{In-beam measurement of the
  hydrogen hyperfine splitting and prospects for antihydrogen spectroscopy},
  Nature Communications 8 (2017) 15749 EP --.
\newline\urlprefix\url{http://dx.doi.org/10.1038/ncomms15749}

\bibitem{BYL00}
Y.~Bylinsky, A.~M. Lombardi, W.~Pirkl, {RFQD}: A decelerating radio frequency
  quadrupole for the {CERN} antiproton facility, Proceedings of the XX
  International Linac Conference C000821 (2000) 554--556.

\bibitem{musashi12}
N.~Kuroda, H.~A. Torii, Y.~Nagata, M.~Shibata, Y.~Enomoto, H.~Imao, Y.~Kanai,
  M.~Hori, H.~Saitoh, H.~Higaki, A.~Mohri, K.~Fujii, C.~H. Kim, Y.~Matsuda,
  K.~Michishio, Y.~Nagashima, M.~Ohtsuka, K.~Tanaka, Y.~Yamazaki,
  \href{https://link.aps.org/doi/10.1103/PhysRevSTAB.15.024702}{Development of
  a monoenergetic ultraslow antiproton beam source for high-precision
  investigation}, Phys. Rev. ST Accel. Beams 15 (2012) 024702.
\newblock \href {http://dx.doi.org/10.1103/PhysRevSTAB.15.024702}
  {\path{doi:10.1103/PhysRevSTAB.15.024702}}.
\newline\urlprefix\url{https://link.aps.org/doi/10.1103/PhysRevSTAB.15.024702}

\bibitem{MOH03}
A.~Mohri, Y.~Yamazaki, \href{http://stacks.iop.org/0295-5075/63/i=2/a=207}{A
  possible new scheme to synthesize antihydrogen and to prepare a polarised
  antihydrogen beam}, EPL (Europhysics Letters) 63~(2) (2003) 207.
\newline\urlprefix\url{http://stacks.iop.org/0295-5075/63/i=2/a=207}

\bibitem{CUSP_Yasu}
Y.~Nagata, Y.~Yamazaki, A novel property of anti-helmholz coils for in-coil
  syntheses of antihydrogen atoms: formation of a focused spin-polarized beam.,
  New J. Phys. 16 ((2014)) 083026.

\bibitem{CUSP_Nagata}
Y.~Nagata, {et al.}, The development of the superconducting double cusp magnet
  for intense antihydrogen beams, Journal of Physics: Conference Series 635 2
  ((2015)) 022062.

\bibitem{ASACUSA_kuroda_nature_comm}
N.~Kuroda, S.~Ulmer, D.~J. Murtagh, S.~Van~Gorp, Y.~Nagata, M.~Diermaier,
  S.~Federmann, M.~Leali, C.~Malbrunot, V.~Mascagna, O.~Massiczek,
  K.~Michishio, T.~Mizutani, A.~Mohri, H.~Nagahama, M.~Ohtsuka, B.~Radics,
  S.~Sakurai, C.~Sauerzopf, K.~Suzuki, M.~Tajima, H.~A. Torii, L.~Venturelli,
  B.~Wuenschek, J.~Zmeskal, N.~Zurlo, H.~Higaki, Y.~Kanai, E.~Lodi~Rizzini,
  Y.~Nagashima, Y.~Matsuda, E.~Widmann, Y.~Yamazaki,
  \href{http://dx.doi.org/10.1038/ncomms4089}{A source of antihydrogen for
  in-flight hyperfine spectroscopy}, Nat Commun 5.
\newline\urlprefix\url{http://dx.doi.org/10.1038/ncomms4089}

\bibitem{Malbrunot20170273}
C.~Malbrunot, C.~Amsler, S.~Arguedas~Cuendis, H.~Breuker, P.~Dupre, M.~Fleck,
  H.~Higaki, Y.~Kanai, B.~Kolbinger, N.~Kuroda, M.~Leali, V.~M{\"a}ckel,
  V.~Mascagna, O.~Massiczek, Y.~Matsuda, Y.~Nagata, M.~C. Simon, H.~Spitzer,
  M.~Tajima, S.~Ulmer, L.~Venturelli, E.~Widmann, M.~Wiesinger, Y.~Yamazaki,
  J.~Zmeskal,
  \href{http://rsta.royalsocietypublishing.org/content/376/2116/20170273}{The
  {ASACUSA} antihydrogen and hydrogen program: results and prospects},
  Philosophical Transactions of the Royal Society of London A: Mathematical,
  Physical and Engineering Sciences 376~(2116).
\newblock \href
  {http://arxiv.org/abs/http://rsta.royalsocietypublishing.org/content/376/2116/20170273.full.pdf}
  {\path{arXiv:http://rsta.royalsocietypublishing.org/content/376/2116/20170273.full.pdf}},
  \href {http://dx.doi.org/10.1098/rsta.2017.0273}
  {\path{doi:10.1098/rsta.2017.0273}}.
\newline\urlprefix\url{http://rsta.royalsocietypublishing.org/content/376/2116/20170273}

\bibitem{Kroyer}
T.~Kroyer, \href{http://cds.cern.ch/record/1098371}{{Design of a spin-flip
  cavity for the measurement of the antihydrogen hyperfine structure}}, Tech.
  Rep. CERN-AB-Note-2008-016, CERN, Geneva (Apr 2008).
\newline\urlprefix\url{http://cds.cern.ch/record/1098371}

\bibitem{Federmann_proceedings}
S.~Federmann, F.~Caspers, E.~Mahner, B.~Juhasz, E.~Widmann, {Design of a 1.42
  GHz Spin-Flip Cavity for Antihydrogen Atoms}, Conf. Proc. C100523 (2010)
  MOPE054.

\bibitem{Silkethesis}
S.~Federmann, \href{http://othes.univie.ac.at/22584/}{A spin-flip cavity for
  microwave spectroscopy of antihydrogen}, dissertation, Universit\"at Wien.
  Fakult\"at f\"ur Physik (2012).
\newline\urlprefix\url{http://othes.univie.ac.at/22584/}

\bibitem{Nagata2016}
Y.~Nagata, N.~Kuroda, M.~Ohtsuka, M.~Leali, E.~Lodi-Rizzini, V.~Mascagna,
  M.~Tajima, H.~Torii, N.~Zurlo, Y.~Matsuda, L.~Venturelli, Y.~Yamazaki, Direct
  detection of antihydrogen atoms using a {BGO} crystal, Nuclear Inst. and
  Methods in Physics Research, A 840 (2016) 153--159.
\newblock \href {http://dx.doi.org/10.1016/j.nima.2016.10.019}
  {\path{doi:10.1016/j.nima.2016.10.019}}.

\bibitem{NAG08}
Y.~Nagata, N.~Kuroda, B.~Kolbinger, M.~Fleck, C.~Malbrunot, V.~Mäckel,
  C.~Sauerzopf, M.~Simon, M.~Tajima, J.~Zmeskal, H.~Breuker, H.~Higaki,
  Y.~Kanai, Y.~Matsuda, S.~Ulmer, L.~Venturelli, E.~Widmann, Y.~Yamazaki,
  \href{http://www.sciencedirect.com/science/article/pii/S0168900218311227}{Monte-carlo
  based performance assessment of {ASACUSA}’s antihydrogen detector}, Nuclear
  Instruments and Methods in Physics Research Section A: Accelerators,
  Spectrometers, Detectors and Associated Equipment 910 (2018) 90 -- 95.
\newblock \href {http://dx.doi.org/https://doi.org/10.1016/j.nima.2018.09.013}
  {\path{doi:https://doi.org/10.1016/j.nima.2018.09.013}}.
\newline\urlprefix\url{http://www.sciencedirect.com/science/article/pii/S0168900218311227}

\bibitem{ClemensNIMA}
C.~Sauerzopf, A.~A. Capon, M.~Diermaier, M.~Fleck, B.~Kolbinger, C.~Malbrunot,
  O.~Massiczek, M.~C. Simon, S.~Vamosi, J.~Zmeskal, E.~Widmann,
  \href{http://www.sciencedirect.com/science/article/pii/S0168900216305630}{Annihilation
  detector for an in-beam spectroscopy apparatus to measure the ground state
  hyperfine splitting of antihydrogen}, Nuclear Instruments and Methods in
  Physics Research Section A: Accelerators, Spectrometers, Detectors and
  Associated Equipment 845 (2017) 579 -- 582, proceedings of the Vienna
  Conference on Instrumentation 2016.
\newblock \href {http://dx.doi.org/https://doi.org/10.1016/j.nima.2016.06.023}
  {\path{doi:https://doi.org/10.1016/j.nima.2016.06.023}}.
\newline\urlprefix\url{http://www.sciencedirect.com/science/article/pii/S0168900216305630}

\bibitem{BerniEXA}
B.~Kolbinger, C.~Amsler, H.~Breuker, M.~Diermaier, P.~Dupr\'e, M.~Fleck,
  A.~Gligorova, H.~Higaki, Y.~Kanai, T.~Kobayashi, M.~Leali, V.~M\"ackel,
  C.~Malbrunot, V.~Mascagna, O.~Massiczek, Y.~Matsuda, D.~Murtagh, Y.~Nagata,
  C.~Sauerzopf, M.~Simon, M.~Tajima, S.~Ulmer, N.~Kuroda, L.~Venturelli,
  E.~Widmann, Y.~Yamazaki, J.~Zmeskal,
  \href{https://doi.org/10.1051/epjconf/201818101003}{Recent developments from
  {ASACUSA} on antihydrogen detection}, EPJ Web of Conferences 181 (2018)
  01003.
\newblock \href {http://dx.doi.org/10.1051/epjconf/201818101003}
  {\path{doi:10.1051/epjconf/201818101003}}.
\newline\urlprefix\url{https://doi.org/10.1051/epjconf/201818101003}

\bibitem{McCullough}
R.~W. McCullough, J.~Geddes, A.~Donnelly, M.~Liehr, M.~P. Hughes, H.~B.
  Gilbody, \href{http://stacks.iop.org/0957-0233/4/i=1/a=013}{A new microwave
  discharge source for reactive atom beams}, Measurement Science and Technology
  4~(1) (1993) 79.
\newline\urlprefix\url{http://stacks.iop.org/0957-0233/4/i=1/a=013}

\bibitem{DIE12}
M.~V. Diermaier, \href{https://othes.univie.ac.at/19893}{Design and
  construction of a monoatomic hydrogen beam}, diploma thesis, Universit\"at
  Wien. Fakult\"at f\"ur Physik (2012).
\newline\urlprefix\url{https://othes.univie.ac.at/19893}

\bibitem{Sextupole_CERN}
P.~A. Thonet, Use of permanent magnets in multiple projects at cern, IEEE
  Transactions on Applied Superconductivity 26~(4) (2016) 1--4.
\newblock \href {http://dx.doi.org/10.1109/TASC.2016.2532361}
  {\path{doi:10.1109/TASC.2016.2532361}}.

\bibitem{WAL82}
J.~T.~M. Walraven, I.~F. Silvera,
  \href{https://doi.org/10.1063/1.1137152}{Helium temperature beam source of
  atomic hydrogen}, Review of Scientific Instruments 53~(8) (1982) 1167--1181.
\newblock \href {http://arxiv.org/abs/https://doi.org/10.1063/1.1137152}
  {\path{arXiv:https://doi.org/10.1063/1.1137152}}, \href
  {http://dx.doi.org/10.1063/1.1137152} {\path{doi:10.1063/1.1137152}}.
\newline\urlprefix\url{https://doi.org/10.1063/1.1137152}

\bibitem{WIE16}
M.~Wiesinger, \href{http://katalog.ub.tuwien.ac.at/AC13480096}{Design and
  implementation of new optics for the atomic hydrogen beam of {ASACUSA}'s
  antihydrogen hyperfine spectroscopy experiment}, diploma thesis, Technische
  Universit{\"a}t Wien (2017).
\newline\urlprefix\url{http://katalog.ub.tuwien.ac.at/AC13480096}

\bibitem{WOL13}
M.~V. Wolf, \href{https://othes.univie.ac.at/30206}{Lock-in based detection
  scheme for a hydrogen beam}, diploma thesis, Universit\"at Wien. Fakult\"at
  f\"ur Physik (2013).
\newline\urlprefix\url{https://othes.univie.ac.at/30206}

\end{thebibliography}


\end{document}